\documentclass[12pt]{article}
\usepackage{array}
\newcolumntype{P}[1]{>{\centering\arraybackslash}p{#1}}
\newcolumntype{M}[1]{>{\centering\arraybackslash}m{#1}}

\usepackage{amsmath,amsthm,amsfonts}
\usepackage{enumerate}
\usepackage{natbib}
\usepackage{url} 

\usepackage{graphicx,psfrag,epsf}

\DeclareMathOperator*{\argmin}{arg\,min}
\usepackage{setspace}
 \DeclareGraphicsExtensions{.eps, .ps}
\usepackage{soul,color}

\usepackage{newfloat}
\usepackage{multirow}
\usepackage{arydshln}

\usepackage{xcolor}

\usepackage{xr}
\makeatletter
\newcommand*{\addFileDependency}[1]{
  \typeout{(#1)}
  \@addtofilelist{#1}
  \IfFileExists{#1}{}{\typeout{No file #1.}}
}
\makeatother

\newcommand*{\myexternaldocument}[1]{%
    \externaldocument{#1}%
    \addFileDependency{#1.tex}%
    \addFileDependency{#1.aux}%
}
\myexternaldocument{supp}

\def\E{\mathbb{E}}

\def\P{\mathbb{P}}
\def\given{\, | \,}

\newtheorem{theorem}{Theorem}[section]

\newtheorem{lemma}[theorem]{Lemma}
\newtheorem{assumption}[theorem]{Assumption}

\newtheorem{condition}[theorem]{Condition}
\newtheorem{proposition}[theorem]{Proposition}
\newtheorem{example}[theorem]{Example}

\renewcommand{\baselinestretch}{0.8}

\newcommand{\blind}{1}

\begin{document}

\def\spacingset#1{\renewcommand{\baselinestretch}%
{#1}\small\normalsize} \spacingset{1}


\if1\blind
{
  \title{\bf Incorporating Auxiliary Variables to Improve the Efficiency of Time-Varying Treatment Effect Estimation}
   \author{Jieru Shi,
    Zhenke Wu, 
    Walter Dempsey \\
    Department of Biostatistics, University of Michigan, USA}
    \date{}
  \maketitle
} \fi

\if0\blind
{
  \bigskip
  \bigskip
  \bigskip
  \begin{center}
    {\LARGE\bf Incorporating Auxiliary Variables to Improve the Efficiency of Time-Varying Treatment Effect Estimation}
\end{center}
  \medskip
} \fi

\bigskip
\begin{abstract}
Contextual sensing and delivery of digital interventions to improve health outcomes have gained significant traction in behavioral and psychiatric studies. Micro-randomized trials (MRTs) are a common experimental design for obtaining data-driven evidence on the effectiveness of digital interventions where each individual is repeatedly randomized to receive treatments over numerous time points. Throughout the study, individual characteristics and contextual factors around randomization are collected, with some prespecified as moderators for assessing time-varying causal effect moderation. However, many additional measurements beyond these moderators often go underutilized. Some of these may influence treatment randomization or known to strongly moderate the treatment effect. Incorporating such auxiliary information into the estimation procedure can reduce chance imbalances and improve asymptotic estimation efficiency. In this work, we propose a method to adjust for auxiliary variables in consistently estimating time-varying intervention effects. The approach can also be extended to include post-treatment auxiliary variables when evaluating lagged treatment effects. Under specific conditions, local efficiency gains are guaranteed. We demonstrate the method's utility through simulation studies and an analysis of data from the Intern Health Study \citep{necamp2020}.
\end{abstract}



\noindent%
{\it Keywords:}  Causal Inference; Asymptotic Efficiency; Micro-randomized Trials; Mobile Health; Moderation Effect; Covariate Adjustment.
\vfill

\newpage
\spacingset{1.8}

\section{Introduction}

The use and development of mobile interventions is experiencing rapid growth. In “just-in-time” mobile interventions \citep{nahum2018}, treatments are provided via mobile devices that are intended to help individuals make healthy decisions ``in the moment'' and thus have a proximal, near-term impact on health outcomes. Micro-randomized trials (MRTs; \cite{klasnja2015}; \cite{dempsey2015}) provide data to guide the development of such mobile interventions~\citep{free2013}, where each participant is sequentially randomized to different treatment options at possibly hundreds to thousands of occasions. 
In an MRT, measurements of individual characteristics, context, and response to treatments are also collected repeatedly, often through passive sensing or self-report. Some of these measurements may be identified in advance as primary moderators of interest when assessing time-varying treatment effect, while the others are referred to as \emph{auxiliary variables}. MRTs admit the definition of a class of causal estimands known as ``causal excursion effects'' \citep{boruvka2018}, and the precise estimation of these effects help inform domain scientists whether to include the treatment in an mHealth intervention package.

The weighted and centered least squares (WCLS) method \citep{boruvka2018} is the state-of-the-art for estimating causal excursion effects. 
It can be framed as a two-stage regression approach. In the first stage, the observed history is trained to predict a proximal or possibly lagged outcome. In the second stage, the debiased outcomes are regressed onto a low-dimensional treatment effect model for interpretable results. 
While this procedure ensures consistent and asymptotically normal estimates of the causal excursion effect, it does not take full advantage of the rich auxiliary information available over time, which can lead to suboptimal precision. In many cases, certain pre-treatment auxiliary variables may be known to strongly moderate the treatment effect, or when studying lagged causal effects, post-treatment auxiliary variables can often be highly predictive. Incorporating these variables into estimation can potentially reduce noise and improve the precision of causal estimates, allowing scientists to make better use of the available data.

Similar issues have been thoroughly explored in the context of single-point treatments in randomized controlled trials (RCTs), particularly when estimating the average treatment effect (ATE). Covariate adjustment is a key technique in this setting, addressing chance imbalances in baseline covariates and maintaining correct type I error rates when covariates are part of the randomization process \citep{kahan2014risks,tackney2023comparison}. Moreover, incorporating ``mean-centered" covariates into regression analyses has been shown to improve asymptotic precision when all treatment-covariate interactions are included in the model 
\citep{lin, ye2022, negi2021revisiting}. In MRTs, auxiliary variables may be predictive of the repeated outcomes and be used to improve the efficiency of estimating the causal excursion effects. However, incorporating these variables presents challenges due to their time-varying nature in the context of sequential randomization studies. A critical question this paper seeks to address is ``How can `mean-centered' time-varying auxiliary variables be incorporated to avoid causal bias and guarantee either  efficiency gains or no efficiency loss?''


\textcolor{black}{We will show in this paper that ``centering'' of a time-varying auxiliary variable is not uniquely defined in our more general context. To see this, consider time-specific mean centering, which naturally extends \cite{lin}'s covariate adjustment to the time-varying setting. In an MRT with many decision points, this choice will lead to a large number of nuisance centering parameters. Moreover, this choice only guarantees consistency for fully marginal causal effects. 
We will show that for many moderation analyses, this choice will lead to causal bias. 
A low-dimensional alternative such as a global weighted mean might be more appealing; however, an incorrect choice will again lead to biased estimates. This motivates our proposal of a practical and theoretically sound framework for incorporating auxiliary variables into the estimating procedure with the goal of improving efficiency while ensuring consistent estimation. }

\paragraph{Main Contributions.} 
\textcolor{black}{In this paper, we propose the Auxiliary-variable Adjusted WCLS (A2-WCLS) method to estimate causal excursion effects in MRTs. Our approach consists of two key components: (1) Pre-treatment auxiliary variable adjustment, where we establish an orthogonality condition that ensures consistent causal parameter estimation while maintaining a low-dimensional estimating equation for nuisance centering parameters, preventing their prohibitive growth with increasing time points. (2) Post-treatment auxiliary variable adjustment, where we show how to incorporate highly predictive post-treatment variables without introducing bias, further enhancing estimation precision. We prove that A2-WCLS guarantees a local efficiency gain when the causal effect model is correctly specified. Extensive simulations demonstrate the significant efficiency gains achieved with this approach. Finally, we apply A2-WCLS to a real-world mobile health study, demonstrating how the improved estimation efficiency allows domain scientist to have more precise evaluation of the magnitude and duration of time-varying treatment effects.
}


The paper is organized as follows: Section \ref{sec:existmeth} reviews current methods for MRTs and covariate adjustment in single-time-point RCTs, highlighting challenges in time-varying settings. Section \ref{sec:estimand} introduces our strategy for incorporating auxiliary variables and outlines its asymptotic properties. Section \ref{sec:theoretical} presents the main theorem on efficiency improvements from proposed auxiliary adjustment. Section \ref{sec:properties} summarizes the statistical properties of the proposed method. Section \ref{sec:sim} includes simulations comparing different estimators and showcasing efficiency gains, while Section \ref{sec:casestudy} demonstrates these gains using data from the Intern Health Study \citep{necamp2020}. The paper concludes in Section \ref{sec:discussion}, with technical proofs and additional results provided in the Supplementary Materials.


\section{Preliminaries}
\label{sec:existmeth}


For a given individual $j$, let $A_{t,j}$ denote the treatment at the $t$-th occasion, where $t = 1, \dots , T; j= 1,\dots,N$. We define \( Y_{t+\Delta,j} \) as the outcome measured at future decision points within a fixed window \( \Delta \geq 1 \). In the special case where \( \Delta = 1 \), \( Y_{t+1,j} \) represents the subsequent proximal response. Individual $j$'s contextual information at the $t$-th treatment occasion is represented by $X_{t,j}$, which may contain summaries of historical context, treatments, or response measurements. For example, before each treatment occasion, the individual might report their current mood status, then the vector $X_{t,j}$ could contain this measurement or, with previous measurements, variation or change in mood. Hereafter, we may omit the subscript $j$ for ease of presentation unless otherwise clarified. The overbar ``$\bar{\bullet}$'' is used to denote a sequence of random variables or realized values up to a specific treatment occasion. For example, $ \bar A_{t-1} =(A_1,...,A_{t-1}$) denotes the sequence of treatments up to and including decision time $t-1$. Information accrued up to treatment occasion $t$ is represented by the history $H_t = (\bar A_{t-1},\bar X_t)$. For simplicity, we present the main results in this paper under a binary treatment $A_{t} \in \{0,1\}$ which follows a known sequence of randomization probabilities $p_t$ that may depend on the complete observed history, denoted $\{p_t = p_t(A_t|H_t)\}_{t=1}^T$.  In the following, random variables or vectors are denoted with uppercase letters; lowercase letters denote their realized values. In addition, we assume that the longitudinal data are independent and identically distributed across $N$ individuals. Note that this assumption would be violated, if, for example, some of the treatments are used to enhance social support between individuals in the study \citep{liao2016}.




\subsection{Causal Excursion Effects and Existing Inferential Methods}
\label{sec:boruvka}



Many treatments are designed to influence an individual in the short-term (or ``proximal'') \citep{heron2010}. To answer questions related to the causal effect of time-varying treatments on the proximal or lagged response, we focus on a class of estimands referred to as ``causal excursion effects". In particular, a causal excursion effect is a function of the decision point~$t$ and a set of moderators~$S_t$, marginalizing over all other observed and unobserved variables \citep{boruvka2018,qian2021}. We provide formal definitions below using potential outcomes~\citep{Rubin, Robins}.

Let~$Y_{t+\Delta} (\bar a_{t+\Delta-1})$ denote the potential outcome for the proximal response under treatment sequence~$\bar a_{t+\Delta-1}$. Let $S_t (\bar a_{t-1})$ denote the potential outcome for a time-varying effect moderator which is a deterministic function of the potential history up to time $t$, $H_t (\bar a_{t-1})$. The $\Delta$-lag causal excursion effect is defined with respect to a reference distribution $\{\pi_{t+u}(a_{t+u} | H_{t+u})\}_{u=1}^{\Delta-1}$, i.e., the treatment randomization probabilities from $t+1$ to $t+\Delta-1$. This generalization contains previous definitions such as the natural lagged effects~\citep{boruvka2018} where $\pi_{t+u} = p_{t+u}$ and deterministic choices such as $a_{t+1:(t+\Delta-1)} = {\bf 0}$~\citep{dempsey2020,qian2021}, where $\pi_{t+ u} = {\bf 1}\{a_{t+ u} = 0\}$ and ${\bf 1}\{\cdot\}$ is the indicator function. Then the $\Delta$-lag causal excursion effect is defined as follows:
\begin{equation}
    \label{eq:causalexcursion_po} 
\beta_{\bf p,\pi}(t+\Delta-1;s) = \E_{{\bf p, \pi}} \left [ Y_{t+\Delta} \left(\bar A_{t-1}, A_t = 1 \right) - Y_{t+\Delta} \left(\bar A_{t-1}, A_t = 0 \right) \given S_t (\bar A_{t-1}) = s\right]. 
\end{equation} 
Equation~\eqref{eq:causalexcursion_po} is defined with respect to both \(\mathbf{p}\) and \(\pi\), the distributions of treatment sequence \(\bar{A}_{t-1}\) and \(A_{t+1:(t+\Delta-1)}\), respectively. We follow common approaches in observational mobile health studies, where analyses are performed marginally over \(\mathbf{p}\) and \(\pi\), such as GEEs \citep{liang1986}. To express the causal excursion effect $\beta_{\bf p, \pi}(t;s)$ in terms of the observed data, we assume positivity, consistency, and sequential ignorability \citep{robins1994,robins1997}:
\begin{assumption}[Identification]
\label{ass:po}\normalfont
~\
  \begin{itemize}
  \item Consistency: 
    $\{Y_{t+\Delta,j} (\bar{A}_{t+\Delta-1,j} ), X_{t,j} (\bar A_{t-1,j}), A_{t,j} (\bar{A}_{t-1,j} )\}  = \{Y_{t+1, j}, X_{t,j}, A_{t,j}\}$, for each $t \leq T$ and $j$, i.e., observed values equal the corresponding potential outcomes;
  \item Positivity: if the joint density~$\P\{H_t = h_t\}$ is greater
    than zero, then~$\P (A_t = a \given H_t = h_t ) > 0$, $a=0,1$;
  \item Sequential ignorability: For each~$t \leq T$, the potential outcomes 
  \newline
  $\{Y_{t+\Delta}(\bar a_t), X_{t+1}(\bar a_t), A_{t+1}(\bar a_t), \dots, Y_{T+1}(\bar a_T)\}$ are independent of $A_{t}$ conditional on the observed history $H_t$.
  \end{itemize}
\end{assumption}

We consider the setting in which the potential outcomes are i.i.d over individuals according to a distribution $\mathcal{P}$. Under Assumption~\ref{ass:po}, the $\Delta$-lag causal excursion effect defined in Equation~\eqref{eq:causalexcursion_po} can be expressed in terms of observable quantities as  \citep{shi2022assessing}:
\begin{equation}
\label{eq:causalexursion}
    \beta_{\bf p,\pi}(t+\Delta-1;s)= \E \big[ \E \left[ W_{t,\Delta-1} Y_{t+\Delta} \mid A_t = 1, H_t \right] - \E \left[ W_{t,\Delta-1} Y_{t+\Delta} \mid A_t = 0, H_t \right] \mid S_t = s \big], 
\end{equation}
where $W_{t, u} = \prod_{s=1}^{u} \pi_{t+s} (A_{t+s} | H_{t+s}) /  p_{t+s} (A_{t+s} | H_{t+s})$. Assuming $\beta_{\bf p,\pi}(t+\Delta;s) = f_t(s)^\top \beta_0^\star$, where $f_t(s) \in \mathbb{R}^q$ is a $q$-dimensional feature vector with a summary of observed information depending only on the effect moderator value $s$ and decision point $t$, the WCLS criterion for \(\Delta\)-lag causal effect is to minimize the following expression \citep{boruvka2018}:
\begin{equation}
\label{eq:mrtstandard}
     \mathbb{P}_N \Big[ \sum_{t=1}^{T-\Delta+1} W_{t,\Delta-1} W_t \times \big( Y_{t+\Delta} - g_t(H_t)^\top \alpha_0 - \left ( A_t - \tilde p_t (1 \mid S_t) \right) f_t (S_t)^\top \beta_0\big)^2 \Big].
\end{equation}
where~$\mathbb{P}_N$ is an operator denoting the sample average, $W_t = \tilde p_t (A_t \mid S_t) / p_t (A_t \mid H_t)$ is a weight where the numerator is an arbitrary function with range $(0,1)$ that only depends on $S_t$, and $g_t(H_t) \in \mathbb{R}^d$ are $d$ control variables chosen to help reduce noise and to construct more powerful test statistics. See \citet{boruvka2018} for more details on the estimand definition and consistency, asymptotic normality, and robustness properties of the WCLS estimation method.

In addition to studying continuous outcomes, there are also many MRTs concerning longitudinal binary outcomes. \cite{qian2021} proposed an estimator of the marginal excursion effect (EMEE) by defining a log relative risk model for the causal excursion effect, and we present a detailed review in Appendix \ref{app:sec_binary}. 


\subsection{Literature Review: Covariate Adjustment}
\label{subsection:checklist}


The RCT methodology literature has extensively examined the use of covariate adjustment to improve the precision of inference on ATEs. Consider, for example, an RCT with a binary randomized treatment denoted $A$, and $N$ subjects in the study are the population of interest. \cite{neyman1923} showed that the difference-in-means estimator is unbiased for the ATE. \cite{fisher1935} proposed to use the ordinary least squares (OLS) adjusted estimator of ATE, which is the estimated coefficient on treatment $A$ in the OLS regression of $Y$ on $\{1, A, X_0\}$. This approach aims to leverage prognostic information in baseline covariates to improve estimation efficiency. However, \cite{freedman2008} demonstrated that in specific scenarios, such adjustment could potentially harm asymptotic precision.
 
\cite{lin} argued that in sufficiently large samples, the issues related to ``adjustment worsening asymptotic precision'' \cite{freedman2008} raised are either minor or easily fixed. In addition, \cite{lin} shows that OLS adjustment with a full set of treatment and covariate interactions improves or does not hurt asymptotic precision, even when the regression model is incorrect. The robust Eicker-Huber–White \citep{eicker1967,huber1967,white2014} variance estimator \citep{freedman2006} is consistent or asymptotically conservative (regardless of whether the interactions are included) in estimating the true asymptotic variance.  More results on the asymptotic precision of treatment effect can be found in papers by \cite{yang2001}, \cite{tsiatis2008}, \cite{vansteelandt2018improving} and \cite{negi2021revisiting}. More complex randomization schemes have led to recent developments in covariate adjustment methods. For example, \cite{su2021} extended the theory to cluster-randomized experiments.
\cite{ye2022} used a model-assisted approach for covariate adjustment and generalized to linear contrast, risk ratios, and odds ratios. Their conclusions are based on an asymptotic theory that provides a clear picture of how covariate-adaptive randomization and regression adjustment alter statistical efficiency.

\textcolor{black}{\cite{guo2021} and \cite{ye2022} both proposed checklists of ideal properties for covariate adjustment methods. As we consider time-varying treatment effect estimation, we build upon their lists of the ideal properties that the auxiliary variable adjustment method should have: (1) valid statistical inference; (2) robust estimation; (3) strong finite-sample performance; (4) wide applicability; and (5) computational simplicity.  In the following sections, we propose and evaluate a general method that satisfies the five ideal properties for performing auxiliary variable adjusted WCLS, termed ``A2-WCLS", for improving efficiency of estimating causal excursion effects in longitudinal data with time-varying treatments.}

\section{Estimand and Inferential Method}
\label{sec:estimand}

\textcolor{black}{
We focus on two key causal estimands in longitudinal settings: the time-specific marginal causal effect and the time-smoothed moderated causal effect. We develop a general estimation framework that incorporates auxiliary variables to improve the estimation of causal excursion effects.
The main theoretical findings on asymptotic efficiency comparisons are summarized in Table \ref{tab:summary}. The results here help answer the earlier question of whether using auxiliary variables is beneficial and how to effectively include them in the estimation process}. To begin, we make a parametric assumption about the causal parameter of interest:


\begin{assumption}
\label{ass:directeffect}
    Assume the causal excursion effect $\beta_{\bf p,\pi}(t+\Delta-1;s) = f_t(s)^\top \beta_0^\star$, where $f_t(s) \in \mathbb{R}^q$ is a feature vector comprised of a $q$-dimensional summary of observed information depending only on state $s$ and decision point $t$.
\end{assumption}
\textcolor{black}{This parametric assumption assumes a correct specification of the causal effect; however, when model misspecification occurs,
we can still interpret the proposed linear form as an $L_2$ projection of the true causal excursion effect onto the space spanned by a $q$-dimensional feature vector $f_t(s)$ that only depends on  $t$ and $s$ \citep{shi2022assessing}. The choice between these interpretations, whether it is a correctly specified causal effect or a projection, reflects a bias-variance trade-off. In practical applications, the projection interpretation ensures a well-defined parameter with practical interest \citep{dempsey2020}.}
\begin{table}[htbp]
\caption{Summary of asymptotic efficiency comparisons under Assumptions \ref{ass:po} and \ref{ass:directeffect} among the unadjusted estimator, WCLS \citep{boruvka2018}, the covariate adjustment method by \cite{lin}, and the proposed A2-WCLS estimators for both time-specific and smoothed estimands. }\label{tab:summary}
\begin{center}
\begin{tabular}{M{1in} M{1.5in}M{1.2in}M{1.9in}}
\hline
Estimator &  Assumptions & \multicolumn{2}{c}{Asymptotic Efficiency Comparison} \\
\hline \\[-0.8em]
\multicolumn{4}{c}{\underline{\it Time-specific marginal causal effect(Section 3.1)}}  \\
\\[-0.8em]
Unadjusted  & - & \multicolumn{2}{l}{\quad\quad\quad\quad Baseline} \\
\\[-0.8em]
  \multirow{2}{*}{WCLS}&- &  \multirow{2}{*}{Lemma 3.2} &{$p_t=0.5$: improved or equal; }\\
 &&&{otherwise: not gauranteed}\\
 \\[-0.8em]
 \cite{lin} & - & Lemma 3.3 & Improved or equal\\ 
\hline \\[-0.8em]
 \multicolumn{4}{c}{\underline{\it Time-smoothed moderated causal effect (Section 3.2)}} \\
 \\[-0.8em]
 WCLS &- & \multicolumn{2}{l}{\quad\quad\quad\quad Baseline}   \\
\\[-0.8em]
\multirow{3}{*}{\centering A2-WCLS} & Condition \ref{con:orthogonality}, Assumptions 4.1, 4.2 & Theorem 4.3 & Improved or equal \\
\\[-0.8em]
 & Condition \ref{con:orthogonality},  Proposition \ref{lemma:beta_1}  & Proposition \ref{lemma:beta_1} & Equal \\
\hline
\end{tabular}
\end{center}
\end{table}


\subsection{Time-Specific Causal Excursion Effect Estimation}
\label{sec:time-specific}


To establish a clear connection with previous research on covariate adjustment, we start with the \emph{time-specific causal excursion effect estimand} with $\Delta = 1$, which shares similarities with the ATE estimand in RCTs. Recall that the ATE estimand is typically defined as the linear difference between the expected outcomes under different treatment allocations, i.e., $\E\big[\E[Y|X,A=1] - \E[Y|X,A=0]\big]$. \textcolor{black}{In the context of MRTs, the time-specific marginal causal effect estimand is defined as $\beta_{\mathbf{p}}(t) = \E\big[\E_{\mathbf{p}}[Y_{t+1}|H_t,A_t = 1]- \E_{\mathbf{p}}[Y_{t+1}|H_t,A_t = 0]\big]$, with the additional time stamp and controlling for the treatment randomization probability, which depends on the observed history.} Moving forward, we drop the subscript ${\mathbf{p}}$ to simplify the notation. The time-specific causal excursion effect is estimated by setting $\beta(t; S_t) = \beta_{0,t}^\star$ in Assumption \ref{ass:directeffect}, with $S_t = \emptyset$, allowing for a non-parametric estimation over time. The treatment randomization probability is denoted as $p_t(A_t | H_t) = p_t$. In this subsection, the phrase ``more efficient" explicitly refers to an estimator of $\beta_{0,t}^\star$ that has a smaller asymptotic variance.

\subsubsection{The Unadjusted and the WCLS Estimator}

\sloppy Analysis using solely the data from time-varying outcomes and treatments, $\{Y_{t+1,j}, A_{t,j}\}_{t=1}^T, j= 1,\dots,N$, without adjusting for auxiliary variables yields the \emph{unadjusted estimator}. The unadjusted estimator $\hat\beta_{0,t}^{\text{U}}$ can be obtained by minimizing the following objective function:
\begin{equation}
\label{eq:unadjusted_nonpar}
     \mathbb{P}_N \Big[ \sum_{t=1}^T \big( Y_{t+1} - \alpha_{0,t}  - \left ( A_t -  p_t \right) \beta_{0,t} \big)^2 \Big].
\end{equation}


To potentially improve the precision of causal parameter estimates using auxiliary variables, we could use the objective function presented in Equation \eqref{eq:mrtstandard} with $\beta(t; S_t) = \beta_{0,t}^\star$ and mean-centered control variables $\tilde g_t(H_t) = g_t(H_t) - \E[g_t(H_t)]$. The WCLS estimator, $\hat\beta_{0,t}^{\text{WCLS}}$, is thus obtained by minimizing the following objective function:
\begin{equation}
\label{eq:wcls_nonpar}
     \mathbb{P}_N \Big[ \sum_{t=1}^T  \big( Y_{t+1} - \alpha_{0,t} - \tilde g_t(H_t)^\top \alpha_{1,t}  - \left ( A_t - p_t \right) \beta_{0,t} \big)^2 \Big].
\end{equation}
Both estimators, $\hat\beta_{0,t}^{\text{U}}$ and $\hat\beta_{0,t}^{\text{WCLS}}$, are consistent and asymptotically normal \citep{boruvka2018}. This raises a natural question: does including the additional term consistently improve precision? If not, under what conditions might the asymptotic variance of the causal parameter estimates increase when using $\hat\beta_{0,t}^{\text{WCLS}}$ instead of $\hat\beta_{0,t}^{\text{U}}$? To address this, we present the following lemma:
\begin{lemma}[\bf{Asymptotic Variance Comparison between $\hat\beta_{0,t}^{\text{WCLS}}$ and $\hat\beta_{0,t}^{\text{U}}$}]
\label{lemma:wcls_u_nonpar}
Denote $\beta_{1,t}^\star = \Sigma(g_t(H_t))^{-1}\E[Y_{t+1}(A_t-p_t)\tilde g_t(H_t)]\in \mathbb{R}^{d}$, where $\Sigma(g_t(H_t)) = \E\big[ \tilde g_t(H_t) \tilde g_t(H_t)^\top\big] \in \mathbb{R}^{d\times d}$.
The difference in the asymptotic variance between the WCLS estimator $\hat\beta_{0,t}^{\text{WCLS}}$ and the unadjusted estimator $\hat\beta_{0,t}^{\text{U}}$ is $- p_t (1-p_t) \alpha_{1,t}^{\star\top} \Sigma(g_t(H_t))  \big(\alpha^\star_{1,t}+2(1-2p_t) \beta^\star_{1,t}\big)$.  
\end{lemma}


Here, $\alpha^\star_{1,t}$ denotes the true coefficient of the variables $\tilde g_t(H_t)$ in Equation \eqref{eq:wcls_nonpar}, while $\beta_{1,t}^\star$ represents their true moderation effect size. The least-squares estimator for $\beta_{1,t}^\star$ is obtained by minimizing the objective function in Equation \eqref{eq:a2wcls_pertime}.
If \( p_t = 1/2 \), the adjustment is neutral or beneficial. However, if \( g_t(H_t) \) is positively correlated with the outcome and strongly moderates the treatment effect (i.e., \( \alpha^\ast_{1,t} > 0 \) and \( |\beta^\ast_{1,t}| \) is large), an adjustment may increase the asymptotic variance when \( \beta^\ast_{1,t} > 0 \) and \( p_t > 1/2 \). Thus, the WCLS criterion in \eqref{eq:wcls_nonpar} does not always guarantee efficiency gains over the unadjusted estimator in \eqref{eq:unadjusted_nonpar}, consistent with \cite{freedman2008}'s findings on covariate adjustment in RCTs.


\subsubsection{Proposed and More Efficient Estimator}
\label{sec:a2wcls_pertime}


To ensure improved estimation efficiency, we propose an alternative adjustment method for time-varying auxiliary variables. This approach integrates \cite{lin}'s idea into the WCLS framework, yielding an estimator for the causal excursion effect obtained by minimizing:
\begin{equation}
\label{eq:a2wcls_pertime}
 \mathbb{P}_N \Big[ \sum_{t=1}^T \big( Y_{t+1} - \alpha_{0,t} - \tilde g_t(H_t)^\top \alpha_{1,t} - \left ( A_t -  p_t  \right) (\beta_{0,t} +  \tilde g_t(H_t)^\top \beta_{1,t}) \big)^2 \Big].
\end{equation}
Equation \eqref{eq:a2wcls_pertime} is a strict extension of the proposal in \cite{lin}; we denote the minimizer as $(\hat\alpha_{0,t},\hat\alpha_{1,t}, \hat\beta_{0,t}^{\text{L}}, \hat \beta_{1,t})$.
Lemma~\ref{theorem:a2wcls_pertime} below establishes the asymptotic efficiency gain of $\hat{\beta}_{0,t}^{\text{L}}$ over \emph{both} the unadjusted $\hat{\beta}_{0,t}^{\text{U}}$ and the WCLS $\hat{\beta}_{0,t}^{\text{WCLS}}$ estimators.

\begin{lemma} 
\label{theorem:a2wcls_pertime}
Suppose Assumptions \ref{ass:po} and \ref{ass:directeffect} hold, and the randomization probability $p_t$ is known. Given invertibility and regularity conditions, let $(\hat\alpha_{0,t},\hat\alpha_{1,t}, \hat\beta_{0,t}^{\text{L}}, \hat \beta_{1,t})$ minimize objective function \eqref{eq:a2wcls_pertime}:
\begin{enumerate}
    \item $\hat\beta_{0,t}^{\text{L}}$ is consistent and asymptotically normal such that $\sqrt{N}(\hat\beta_{0,t}^{\text{L}} - \beta_{0,t}^\star) \rightarrow \mathcal{N}(0,Q_t^{-1}\Sigma_t^{\text{L}} Q_t^{-1})$, where $Q_t, \Sigma_t^{\text{L}}$ are defined in Appendix \ref{app:a2wcls_pertime};
    \item  $\hat\beta_{0,t}^{\text{L}}$ is at least as efficient as $\hat\beta_{0,t}^{\text{U}}$ described in \eqref{eq:unadjusted_nonpar}, and the asymptotic efficiency gain is
    $\frac{\left(\alpha^\star_{1,t}+(1-2p_t) \beta^\star_{1,t}\right)^\top \Sigma(g_t(H_t)) \left(\alpha^\star_{1,t}+(1-2p_t) \beta^\star_{1,t}\right)}{p_t (1-p_t)} + \beta_{1,t}^{\star\top}\Sigma(g_t(H_t))\beta^\star_{1,t}$;
    \item $\hat\beta_{0,t}^{\text{L}}$ is at least as efficient as $\hat\beta_{0,t}^{\text{WCLS}}$ described in \eqref{eq:wcls_nonpar}, and the asymptotic efficiency gain is $\frac{1-3p_t + 3p_t^2}{p_t (1-p_t)}  \beta_{1,t}^{\star\top}\Sigma(g_t(H_t))\beta^\star_{1,t}$.
\end{enumerate}
\end{lemma}

Lemma~\ref{theorem:a2wcls_pertime} demonstrates that incorporating mean-centered auxiliary variables in the estimation of the time-specific marginal treatment effect using the objective function in Equation \eqref{eq:a2wcls_pertime} guarantees an efficiency improvement over both the unadjusted and WCLS estimators. 
A detailed proof of Lemma \ref{theorem:a2wcls_pertime} is available in Appendix \ref{app:a2wcls_pertime}. \textcolor{black}{Notably, this efficiency comparison extends beyond marginal effect to also cover moderation effect estimators with \textit{categorical} moderators, i.e., when $\beta(t; S_t) = f_t(S_t)^\top \beta_{0,t}^\star$ and $f_t(S_t)$ represents categorical variables. 
A detailed discussion and proof can be found in Appendix \ref{app:remark3.4}.}




\subsection{Auxiliary Variable Adjusted WCLS (A2-WCLS) Estimation}
\label{sec:time-smoothed}

\textcolor{black}{While estimating equation \eqref{eq:a2wcls_pertime} ensures an efficiency gain, estimating centering means for $g_t(H_t)$ at each time point separately introduces a substantial number of nuisance parameters, which can be impractical given the often large time horizons in MRTs. In addition, centering the auxiliary variable at its mean is not a one-size-fits-all remedy. For example, when we are interested in moderated treatment effects (i.e., $S_t \neq \emptyset$), centering by time-specific means can introduce bias. Finally, in longitudinal studies, lagged causal effects are sometimes of primary interest, but incorporating post-treatment variables requires careful handling to avoid bias. This calls for new methodological approaches. To tackle these challenges, we propose a general estimation framework that incorporates both pre- and post-treatment auxiliary variables when assessing time-varying effect moderation, helping to reduce chance imbalance and improve estimation precision.
}

Here we consider auxiliary variable adjusted estimators of the causal excursion effect, with a \emph{primary moderator} of interest $f_t(S_t)\in \mathbb{R}^q $ and a smoothed function of the causal excursion effect model $f_t(S_t)^\top \beta_0$ as mentioned in Assumption \ref{ass:directeffect}.  Previously in Section \ref{sec:boruvka}, we reviewed the estimator $\hat \beta_0^{\text{WCLS}}$ obtained from the WCLS criterion outlined in Equation \eqref{eq:mrtstandard} is consistent and asymptotically normal \citep{boruvka2018}. 
Our goal is to develop a practical procedure for integrating auxiliary information into the estimation process and to evaluate its asymptotic efficiency compared to that of $\hat \beta_0^{\text{WCLS}}$. In the context of moderation effect analysis, ``greater efficiency'' specifically refers to a reduction in the asymptotic variance of the estimated moderated effect for a given value of $S_t = s$, i.e., a smaller asymptotic $\text{Var}(f_t(s)^\top \hat\beta_0)$.



\subsubsection{Pre-treatment Auxiliary Variable Adjustment}
\label{subsec:A2}

Let $Z_t \in \mathbb{R}^p$ represent a set of $p$-dimensional \emph{auxiliary variables} observed before treatment allocation $A_t$, and $Z_t \cap f_t(S_t) = \emptyset$. For clarity, we set \(\Delta = 1\) in this subsection without loss of generality.
An auxiliary variable adjusted estimation method using a centering function $\mu_t(S_t) \in \mathbb{R}^p$ is developed below. Under Assumption \ref{ass:directeffect}, the Auxiliary-variable Adjusted Weighted and Centered Least Square (A2-WCLS) criterion is to minimize the following:
\begin{equation}
\label{eq:a2wcls}
 \mathbb{P}_N \Big[ \sum_{t=1}^T W_t \times \Big( Y_{t+1} - g_t(H_t)^\top \alpha - \big( A_t - \tilde p_t (1 \mid S_t) \big) \big(f_t (S_t)^\top \beta_0 + (Z_t - \mu_t(S_t))^\top \beta_1 \big) \Big)^2 \Big].
\end{equation}
The resulting minimizer is denoted as $\hat{\beta}_0^{\text{A2}}$, our A2-WCLS estimator. We can also consider including \(Z_t\) as a control variable in the \(g_t(H_t)^\top \alpha\) model. However, since \(g_t(H_t)^\top \alpha\) serves as a working model for \(\E[W_t Y_{t+1} | H_t]\), it is expected to capture sufficient information. To avoid redundancy, we assume \(Z_t \subseteq g_t(H_t)\).

An appropriate choice of the centering function $\mu_t(S_t)$ is key to obtaining a consistent estimation of the moderated causal effect. 
\textcolor{black}{ We must ensure that including the additional interaction term does not bias estimation of $\beta_0$. That is, the minimizers of two objective functions, one in Equation \eqref{eq:a2wcls} and the other in Equation \eqref{eq:mrtstandard}, with and without auxiliary variable adjustment, should both converge asymptotically to the same target value $\beta_0^\star$. 
This can be formulated as a set of linear constraints on the centering function $\mu_t(S_t)$: }



\begin{condition}[\textbf{The orthogonality condition}]
\label{con:orthogonality}
For the $i$-th component of $Z_t \in \mathbb{R}^p$, denoted  $Z^i_t \in \mathbb{R}$, $i=1, \ldots, p$, the associated centering function, denoted~$\mu_t^i (S_t)$, must satisfy the following orthogonality condition with respect to $f_t(S_t) \in \mathbb{R}^q$:
\begin{equation}
\label{eq:orthogonality}
        \E \Big[ \sum_{t=1}^T \tilde p_t (1| S_t) \big(1 - \tilde p_t (1| S_t)\big) \big(Z^i_t-\mu^i_t(S_t)\big) f_t(S_t) \Big] = \mathbf{0}_{q \times 1}.
\end{equation}
where~$\tilde p_t(1|S_t) = \tilde p_t(A_t=1|S_t)$ is the numerator of the weight~$W_t$ that is an arbitrary function with range $(0,1)$ depending only on $S_t$.
\end{condition}


\textcolor{black}{Recognizing that \(\beta_1\) is a nuisance parameter in Equation \eqref{eq:a2wcls} when estimating \(\beta_0\), we draw on the Neyman orthogonality of estimating equations to establish the above condition \citep{neyman1979, chernozhukov2018double}. 
Condition~\ref{con:orthogonality} leads to a low-dimensional centering function \( \mu^{i}_t(S_t) \) that applies uniformly across all time points. This is our key technical contribution, showing how to incorporate time-varying auxiliary variables without compromising the statistical consistency of the estimator.
We illustrate the utility of Condition \ref{con:orthogonality} by considering an initial, simple example:}




\begin{example}[\textbf{Fully marginal causal effect estimation}] 
\label{example:est}
\normalfont
    \textcolor{black}{Consider estimating the fully marginal causal excursion effect $\beta^{\star}_0$, where the primary moderator set is null (i.e., $S_t = \emptyset$ and $f_t(S_t) = 1$). Our goal is to incorporate auxiliary variable $Z_t \in \mathbb{R}$ to improve the estimation. Here we compare three centering function choices: (a) $\mu_t =\bar Z_t$, ($t=1,2,\dots,T$), (b) $\mu_t=\mu =\bar Z$, and (c) $\mu_t=\mu = \frac{\sum_{t=1}^T \tilde p_t (1| S_t) (1 - \tilde p_t (1| S_t))Z_t}{\sum_{t=1}^T \tilde p_t (1| S_t) (1 - \tilde p_t (1| S_t))}$. Option (a) estimates the time-specific mean $\E[Z_t]$ following the approach introduced by \cite{lin}, indicating a correct model specification. It satisfies Equation \eqref{eq:orthogonality}, but involves $T$ estimates, which can be cumbersome when $T$ is large. Option (b), the global mean of the auxiliary variable $\{Z_t\}_{t=1}^T$, is a scalar centering function but can lead to inconsistent estimates of $\beta_0^\star$ since it does not satisfy Equation \eqref{eq:orthogonality}.
    Please refer to Appendix \ref{app:sec:centerby_mean} for an empirical illustration of the biased estimate. Option (c) is a scalar centering function satisfying Equation \eqref{eq:orthogonality}, representing a weighted average of $\{Z_t\}_{t=1}^T$. In conclusion, option (c) has the desirable characteristics we seek — a low-dimensional centering function that is uniformly applicable across all time points and consistently estimates the causal parameter $\beta_0^\star$.} 
\end{example}



\textcolor{black}{This example shows that the consistency of $\hat\beta_0$ does not depend on the specific choice of $\mu_t^i(S_t)$, as long as it satisfies Condition \ref{con:orthogonality}. This flexibility allows \( \mu_t^i(S_t) \) to extend beyond the standard conditional means \( \E[Z_t^i|S_t] \), which are often difficult to estimate consistently, even in moderate dimensions, and increase in number with \( T \). Equation \eqref{eq:orthogonality}, forming a system of \( q \) linear equations, offers a simpler alternative by suggesting that \( \mu_t^i(S_t) \) can be defined as a time-smoothed linear projection of the auxiliary variable \( Z^i_t \) onto the subspace defined by \( f_t(S_t) \in \mathbb{R}^q \). This reduces the nuisance parameters from \( T \) to the \( q \) coefficients of the linear model. Since \( q \) is generally small to maintain interpretability, this approach avoids the prohibitive growth of nuisance parameters with \( T \) while ensuring consistent estimation of the causal parameter of interest.
}



To illustrate the proposal more concretely, we can express the centering function as $\mu^i_t(S_t; \theta^i) = f_t(S_t)^\top \theta^i$, where $i \in \{1,\dots,p\}$, and $\hat\theta^i \in \mathbb{R}^q$ can be computed by solving Equation \eqref{eq:orthogonality}. Let 
$\hat\Theta = (\hat{\theta}^1,\dots,\hat{\theta}^p)\in \mathbb{R}^{q \times p}$, we can construct an easy-to-implement A2-WCLS criterion as follows:
\begin{equation}
\label{eq:a2wcls_working}
 \mathbb{P}_N \bigg[ \sum_{t=1}^T W_t \times \Big( Y_{t+1} - g_t(H_t)^\top \alpha - \big ( A_t - \tilde p_t (1 \mid S_t) \big) \big(f_t (S_t)^\top \beta_0 +  (Z_t - \hat\Theta^\top f_t(S_t) )^\top \beta_1\big) \Big)^2 \bigg].
\end{equation}

The following lemma describes the asymptotic properties of the A2-WCLS estimator, allowing us to compare its estimation precision with the existing WCLS estimator.

\begin{lemma}
\label{theorem:a2wcls_general}
Suppose Assumptions \ref{ass:po} and \ref{ass:directeffect} hold, centering function $\{\mu_t(S_t)\}_{t=1}^T$ satisfies the Orthogonality Condition \ref{con:orthogonality}, and that the randomization probability $p_t(A_t=1 | H_t)$ is known. Let $(\hat\alpha, \hat\beta_0^{\text{A2}}, \hat \beta_1)$ minimize objective function (\ref{eq:a2wcls}). Under regularity conditions, $\hat\beta_0^{\text{A2}}$ is consistent and asymptotically normal such that $\sqrt{N}(\hat\beta_0^{\text{A2}}-\beta_0^\star) \rightarrow \mathcal{N}(0,Q^{-1}\Sigma^{\text{A2}} Q^{-1})$, where $Q, \Sigma^{\text{A2}}$ are defined in Appendix \ref{app:a2-wcls_asymptotic}.
\end{lemma} 

Here we consider $T$ as a fixed value, with the sample size $N$ approaching infinity. In some cases, MRTs may involve small sample sizes compared to the scale of total time points $T$. Our simulations in Section \ref{sec:sim} addressed this by applying a small sample correction \citep{mancl2001} to the robust variance estimator, which yielded strong empirical performance for small sample sizes. 


\subsubsection{Post-Treatment Auxiliary Variable Adjustment}

There is increasing interest in examining how current treatments affect lagged outcomes (\(\Delta > 1\)). In this context, incorporating auxiliary variables follows similar principles as with proximal outcomes but provides a broader set of options. In particular, \emph{post-treatment auxiliary variables}, which are often highly predictive, can be included to help reduce noise and improve the precision of causal parameter estimates. 

\textcolor{black}{All variables included in Equation \eqref{eq:mrtstandard}, as indicated by the notation, must be observed prior to the treatment allocation at time $t$ ($A_t$). Motivated by the efficient influence function of $\beta_0$ in \cite{murphy2001marginal}, we propose to improve the estimator in \eqref{eq:mrtstandard} by subtracting its projection onto the score functions of the nuisance parameters \citep{bickel1993efficient,robins1999testing}, specifically the treatment randomization probabilities between \( t+1 \) and \( t+\Delta-1 \). This adjustment has the potential to enhance efficiency; however, it inherently incorporates post-treatment variables, which, if not carefully accounted for, may introduce bias, as demonstrated in Appendix \ref{app:lag-outcome-bias}. }


\textcolor{black}{To address this challenge, we introduce the following A2-WCLS criterion for lagged outcomes, which effectively incorporates post-treatment auxiliary variables to enhance asymptotic efficiency, while mitigating bias from post-treatment adjustment. Let $l_{t+u}(A_{t+u}, H_{t+u})$ be a working model for $\E[W_{t+u,\Delta -1-u} Y_{t+\Delta}|A_{t+u}, H_{t+u}],~0<u \leq \Delta-1$. Assuming \(\pi_{t+u} = p_{t+u}\), it follows that the weights simplify to \(W_{t+u} = 1\) for all \(u\). We define the centering functions as $\mu^l_{t+u}(H_{t+u})= \E[l_{t+u}(A_{t+u},H_{t+u})|H_{t+u}]= \sum_{a_{t+u}} p_{t+u}(a_{t+u}|H_{t+u})l_{t+u}(a_{t+u},H_{t+u})$. The A2-WCLS estimator for the \(\Delta\)-lag causal effect, \(\hat\beta_0^{\text{A2}}\), is obtained by minimizing:}
{\small
\begin{equation}
\label{eq:a2-wcls-lag}
\begin{split}
    \mathbb{P}_N \Big[ \sum_{t=1}^{T-\Delta+1} & W_t \Big( Y_{t+\Delta} - \sum_{u=1}^{\Delta-1} \big(l_{t+u}(A_{t+u}, H_{t+u}) - \mu^l_{t+u}(H_{t+u})\big) -\\   &~~~~~~~~~~~~~~~~~~~ \mu^l_{t}(H_{t}) -\big ( A_t - \tilde p_t (1 \mid S_t) \big)\big(f_t (S_t)^\top \beta_0 + (Z_t - \mu_t(S_t))^\top \beta_1 \big) \Big)^2 \Big].
\end{split}
\end{equation}}
\textcolor{black}{The objective function \eqref{eq:a2-wcls-lag} preserves the robust estimation property of WCLS, ensuring a consistent estimate of $\beta_0$ even if \( l_{t+u}(A_{t+u}, H_{t+u}) \) is misspecified \citep{boruvka2018}. In the simple case where \(\Delta = 2\), we can specify \( l_{t+u}(A_{t+u}, H_{t+u}) \) as a linear working model \( (A_{t+u} - p_{t+u})(\alpha_{u,1} + f_{t+u}(S_{t+u})^\top\alpha_{u,2}) + f_{t+u}(S_{t+u})^\top\alpha_{u,3} \). This leads to the centering function \(\mu^l_{t+u}(H_{t+u}) = f_{t+u}(S_{t+u})^\top \alpha_{u,3}\). As a result, Equation \eqref{eq:a2-wcls-lag} can be expressed as:
\begin{equation}
\label{eq:a2-wcls-lag2}
    \begin{split}
    \mathbb{P}_N \Big[ \sum_{t=1}^{T-\Delta+ 1} & W_t \Big( Y_{t+\Delta} - \big((A_{t+1} - p_{t+1})(\alpha_{1,1} + f_{t+1}(S_{t+1})^\top\alpha_{1,2}) \big)- \\   &~~~~~~~~~ f_{t}(S_{t})^\top \alpha_{0,3} -\big ( A_t - \tilde p_t (1 \mid S_t) \big)\big(f_t (S_t)^\top \beta_0 + (Z_t - \mu_t(S_t))^\top \beta_1 \big) \Big)^2 \Big].
\end{split}
\end{equation}
This linear working model simplifies the inference procedure, as the estimation of nuisance parameters \( (\alpha_{u,1}, \alpha_{u,2}) \in \mathbb{R}^{q+1} \) does not introduce additional variance into the estimation of the causal parameter \( \beta_0 \). The following lemma shows the asymptotic properties of the A2-WCLS estimator for the \(\Delta\)-lag causal effect. Proofs can be found in Appendix \ref{app:lag-outcome}.
\begin{lemma}
\label{lemma:lag-effect}
    Under the conditions of Lemma \ref{theorem:a2wcls_general} and the reference treatment probability satisfies \( \pi_{t+u} = p_{t+u} \) for all \( 0 < u \leq \Delta - 1 \). Assuming $\beta_{\bf p,\pi}(t+\Delta-1;s) = f_t(s)^\top \beta_0^\star$, and let $\hat\beta_0^{\text{A2}}$ minimize objective function (\ref{eq:a2-wcls-lag}). Under regularity conditions, $\hat\beta_0^{\text{A2}}$ is consistent and asymptotically normal such that $\sqrt{N}(\hat\beta_0^{\text{A2}}-\beta_0^\star) \rightarrow \mathcal{N}(0,Q^{-1}\Sigma_l^{\text{A2}} Q^{-1})$, where $Q, \Sigma_l^{\text{A2}}$ are defined in Appendix \ref{app:lag-outcome}.
\end{lemma}
In the more general case where \( \pi_{t+u} \neq p_{t+u} \), the estimating equation becomes more complex, and we defer it to Appendix \ref{app:lag-outcome}. }

\section{Theoretical Results}
\label{sec:theoretical}


\subsection{Efficiency Comparison of the Proposed Estimator}

We have demonstrated the sufficient condition for obtaining consistent estimates when incorporating auxiliary variables. This section explores the estimation efficiency for time-varying causal effect moderation. Since treatments, moderators, and outcomes exhibit temporal dependence, and the estimators for moderated effects are smoothed functions over time, a global efficiency gain, such as that shown in Lemma \ref{theorem:a2wcls_pertime} for time-specific causal effects or in the theorem from \cite{lin}, is no longer guaranteed. However, local efficiency gains can be achieved under additional assumptions specific to modeling time-varying moderated causal effects.


\begin{assumption}[Correct causal model specification]
\label{con:sufficient_1} 
    \sloppy The true causal effect model $\beta_{\bf p,\pi}(t+\Delta-1;H_t)$), $\Delta \geq 1$ is correctly specified in $H_t$ as a linear model. 
\end{assumption}
\sloppy Here, we do not impose specific distribution assumptions on potential outcomes $\{Y_{t+\Delta}(\bar A_{t-1}, a)\}, a \in \{0,1\}$ and $\Delta \geq 1$, nor require the true data-generating model $\E[Y_{t+\Delta}|H_t, A_t]$ to be correctly specified. We only require the difference in conditional outcomes satisfy $\beta(t+\Delta-1;H_t)= \E[W_{t,\Delta - 1}Y_{t+\Delta}|H_t, 1] - \E[W_{t,\Delta - 1}Y_{t+\Delta}|H_t, 0] = f_t(S_t)^\top \beta^\star_0 + (Z_t - \mu_t(S_t))^\top \beta^\star_1$.

\begin{assumption}[Exogenous residuals]
\label{con:sufficient_2} The residuals in the outcomes and treatments, not accounted for by the observed history, is conditionally independent of the future observations given the complete history.
\end{assumption}


This assumption implies that the residuals do not carry any additional information beyond that which is already contained in the observed history. This is fundamental in many statistical learning models, particularly in Markov Decision Processes (MDPs), where the current state determines future states, independent of any unobserved information. Based on the assumptions outlined above, we can state the following theorem:

\begin{theorem}
\label{th:efficiency} For estimating the \(\Delta\)-lag causal excursion effect, A2-WCLS offers the following efficiency properties compared to the WCLS estimator \citep{boruvka2018}:
\begin{enumerate}
    \item Under the conditions of Lemma \ref{theorem:a2wcls_general} and Assumptions \ref{con:sufficient_1} and \ref{con:sufficient_2}, incorporating auxiliary variables via Equation \eqref{eq:a2wcls} with a known centering function $\mu_t(S_t)$ guarantees an asymptotic efficiency improvement over the WCLS estimator in estimating the moderated causal effect for a given \( S_t = s \), by \( f_t(s)^\top Q^{-1} \Sigma Q^{-1}f_t(s) \), where $Q, \Sigma$ are defined in Appendix \ref{app:varreduction}.
    \item Under the conditions of Lemma \ref{lemma:lag-effect} and Assumptions \ref{con:sufficient_1}, if \( \{l_{t+u}(H_{t+u}, A_{t+u})\}_{u=0}^{\Delta-1} \) are correctly specified, incorporating post-treatment auxiliary variables via Equation \eqref{eq:a2-wcls-lag} achieves semiparametric efficiency for estimating the $\Delta$-lag moderated causal effect. The derivative of Equation \eqref{eq:a2-wcls-lag} with respect to \(\beta_0\) is the efficient influence function for \(\beta_0\).
\end{enumerate}
\end{theorem}
\textcolor{black}{
Although an efficiency gain is not generally guaranteed when the causal effect models are misspecified, the following proposition shows that under certain conditions, adjusting for auxiliary variables using the proposed A2-WCLS estimating equation ensures that the asymptotic variance of the causal parameter estimates does not increase:
\begin{proposition}
\label{lemma:beta_1}
If the auxiliary variables \( Z_t \) is conditionally mean independent of the causal excursion effect given \( S_t \), then incorporating them via A2-WCLS yields an estimator with the same asymptotic variance as the standard WCLS estimator \citep{boruvka2018}. 
\end{proposition}
Recall from Equation \eqref{eq:causalexursion} that $\beta(t; S_t, Z_t)$ is defined as $\E\big[\E[Y_{t+1}|H_t, 1] - \E[Y_{t+1}|H_t, 0] \mid S_t, Z_t\big]$. A straightforward derivation shows that this condition is equivalent to $\beta_1^\star = 0$. To provide context, when selecting auxiliary variables for adjustment, there is a risk of including irrelevant ones. However, this proposition shows that even if non-effect moderators are mistakenly included, the asymptotic variance of the estimator will not be inflated using the A2-WCLS estimating equation.}
 
\subsection{Estimating the centering functions}


\textcolor{black}{Theorem~\ref{th:efficiency} assumes known centering functions $\mu_t(S_t)$ and conditional mean functions $l_{t+u}(H_{t+u}, A_{t+u})$; however, in practice, these functions may need to be estimated from observed data, and analysis should account for this extra variance to ensure proper uncertainty quantification. A ``stacked estimating equations'' approach, a type of M-estimator \citep{carroll2006}, is adopted to construct a robust asymptotic variance estimator. We parameterize the working models for \(\mu_t(S_t)\) and \(l_{t+u}(H_{t+u}, A_{t+u})\) with \(\theta_{\mu}\) and \(\theta_{l}\), respectively, and define \(\Theta = (\theta_{\mu}, \theta_{l})\). The corresponding estimating equation is denoted as \(U(\Theta)\). The estimating equation for \( \beta_0 \), denoted as \( U_{\beta_0}(\beta_0; \hat\Theta, \hat\beta_1) \), is the derivative of the proposed A2-WCLS criterion in Equation \eqref{eq:a2wcls} with respect to $\beta_0$:
\begin{equation}
\label{eq:sim-eq}
\begin{split}
    &U_{\beta_0}(\beta_0; \hat\Theta, \hat\beta_1) = \P_N \Big[ \sum_{t=1}^T W_t \Big(Y_{t+\Delta} -\sum_{u=1}^{\Delta-1} \big(l_{t+u}(A_{t+u}, H_{t+u}; \hat\theta_l) - \mu^l_{t+u}(H_{t+u}; \hat\theta_l)\big)  -\\
    &~~~- \mu_t^{l}(H_t;\hat\theta_l) -( A_t - \tilde p_t(1|S_t)) \big(f_t (S_t)^\top \beta_0 + (Z_t - \mu_t(S_t,\hat\theta_{\mu}))^\top \hat\beta_1 \big)\Big)( A_t - \tilde p_t)f_t(S_t)\Big],
\end{split}
\end{equation} 
where $\hat\beta_1$ is obtained by minimizing Equation \eqref{eq:a2wcls}, and $\hat\Theta$ is obtained from solving $U(\Theta) = 0$.} 
By applying the stacked estimating equations approach, we establish the following asymptotic property for the causal parameter estimate \( \hat\beta^{\text{A2}}_0 \):
\begin{lemma}
\label{lemma:adjusted_sandwich}
Under the conditions of Lemma \ref{theorem:a2wcls_general} or Lemma \ref{lemma:lag-effect}, if we incorporate auxiliary variables via A2-WCLS with estimated functions \( \mu_t(S_t,\hat\theta_{\mu}) \) and \(l_{t+u}(A_{t+u}, H_{t+u}; \hat\theta_l) \), where $\hat\Theta$ is obtained from solving $U(\Theta) = 0$, then \( \hat\beta^{\text{A2}}_0 \) is consistent and asymptotically normal such that $\sqrt{N}(\hat\beta_0^{\text{A2}}-\beta_0^\star) \rightarrow \mathcal{N}(0,Q^{-1}\tilde\Sigma^{\text{A2}} Q^{-1})$, where \( Q \) is as defined in Theorem \ref{th:efficiency}, and \( \tilde\Sigma^{\text{A2}} = \E[\tilde U_{\beta_0}(\hat\beta_0;\hat \Theta, \hat\beta_1)^\top\tilde U_{\beta_0}(\hat\beta_0;\hat\Theta, \hat\beta_1)] \), in which $\tilde U_{\beta_0}(\hat\beta_0;\hat \Theta, \hat\beta_1)$ is obtained as follows:
\begin{equation}
\label{eq:adjusted}
    \tilde U_{\beta_0}(\hat\beta_0;\hat \Theta, \hat\beta_1) = U_{\beta_0}(\hat\beta_0;\hat \Theta, \hat\beta_1) - U(\hat\Theta)\Big(\E\Big[\frac{\partial U(\Theta)}{\partial \Theta} \big\vert_{\Theta = \hat \Theta} \Big]\Big)^{-1}\E\Big[\frac{\partial U_{\beta_0}(\hat\beta_0;\Theta, \hat\beta_1)}{\partial \Theta}\big\vert_{\Theta = \hat \Theta} \Big]^\top.
\end{equation}
\end{lemma}

While the efficiency improvement may be smaller compared to Theorem \ref{th:efficiency}, our proposed method remains robust to the estimated centering functions. We see empirical finite-sample efficiency gains in Section \ref{sec:sim}, and provide a further discussion in Appendix \ref{app:sec:finite_super}.

\section{Properties}
\label{sec:properties}

In this section, we return to the five desired properties in the checklist discussed in Section \ref{subsection:checklist} and demonstrate that A2-WCLS satisfies all of them.

 \textbf{Valid statistical inference.} 
    As demonstrated in Lemma \ref{theorem:a2wcls_general}, \ref{lemma:lag-effect} and Theorem \ref{th:efficiency}, our approach ensures consistent and asymptotically normal estimation of the causal parameter. Moreover, under Assumption \ref{con:sufficient_1} and \ref{con:sufficient_2}, A2-WCLS has theoretical guarantees for achieving an efficiency gain relative to the original WCLS method.

 \textbf{Robust estimation.}
    (1) 
    Proposition \ref{lemma:beta_1} indicates that the inclusion of non-effect-moderator auxiliary variables in the causal effect model does not compromise the precision of the estimation of the parameter of interest. (2) The A2-WCLS method shares the same property as WCLS in that if the true randomization probability $p_t(A_t|H_t)$ is known, a consistent estimator can be obtained even when the nuisance function $g_t(H_t)^\top \alpha$ and $l_{t+u}(H_{t+u},A_{t+u})$ are misspecified \citep{boruvka2018}. Robust standard errors are used to account for model misspecification and heteroskedasticity.

\textbf{Strong finite-sample performance.}
Our conclusions are derived based on asymptotic properties of the estimators. 
However, our proposed method can still provide reliable performance in a finite sample with a small sample correction \citep{mancl2001} applied to the variance estimator, which is elaborated in Appendix \ref{app:ssa}. 

 
 \textbf{Wide applicability.} 
 The proposed method is a rigorous longitudinal extension of the existing literature on covariate adjustment that can be applied to study designs with $T \geq 1$, as discussed in Lemma \ref{theorem:a2wcls_pertime} and Theorem \ref{th:efficiency}. When $T=1$, the adjustment of the auxiliary variables is based on baseline information, subsuming the covariate adjustment techniques discussed in \cite{lin} and \cite{ye2022} as special cases.

 \textbf{Computational simplicity.}
    Using linear working models $\Theta^\top f_t(S_t)$ for the centering function simplifies computation, and can be easily implemented using existing routines found in standard statistical software packages. See the reproducible code and documentations provided with this paper.
    

\section{Simulation}
\label{sec:sim}

Here,  the empirical performance of the proposed A2-WCLS method is examined through simulation. Results verify the theory that the A2-WCLS method can significantly improve the efficiency of causal effect estimation, even when accounting for the additional variance introduced by estimating the centering function from data.



\subsection{Simulation Setup}


The proposed method is empirically evaluated via simulation, where the data generation model is based on \cite{boruvka2018} with minor adjustments made to demonstrate the improvement of our proposed method over the WCLS approach.
Consider an MRT with known randomization probability and observation vector being a single state variable $Z_t \in \{-1,1\}$ with probability $0.5$ taking each value at each decision time $t$,  Let
\begin{equation}
\label{eq:generativemodel}
\begin{split}
    Y_{t+2,j} & = 0.2 Z_{t+1,j} + \big(-0.2  +  0.8 (Z_{t+1,j} - \E[Z_{t+1,j}])\big)\times (A_{t+1,j} -p_{t+1}(1|H_{t+1,j}))+ \\
    &~~~~~~~~~~~~~~~~~~~~~~~~~~~\big(\beta_0  +  \beta_1 (Z_{t,j} - \E[Z_{t,j}])\big)\times (A_{t,j} -p_t(1|H_{t,j})) +\epsilon_{t,j}.
\end{split}
\end{equation}
The randomization probability is set to $p_t(1|H_t) = \text{expit}(\eta_1 A_{t-1}+\eta_2 Z_t)$ where $(\eta_1,\eta_2) = (-0.8,  0.8)$ and $\text{expit}(x)=(1+\exp(-x))^{-1}$. 
The independent error term satisfies $\epsilon_{t} \sim \mathcal{N}(0,1)$ with $\text{Corr}(\epsilon_u, \epsilon_t) = 0.5^{|u-t|/2}$. We set $\beta_{0}=-0.1$, and $\beta_{1} \in \{0.2,0.5,0.8\}$ to represent a small, medium, and large lagged moderation effect.

\subsection{Simulation Results}
\label{sec:sim_2}

\textcolor{black}{In the main simulation experiment, the lag $\Delta = 2$ marginal causal effect is estimated, which is constant in time and is given by $\beta_0^\star =  -0.1$. Results are reported with $N = 250$ individuals and $T = 30$ using the following estimation methods:}

\noindent{\bf Estimation Method I: WCLS}. The WCLS method \citep{boruvka2018} is applied using a fully marginal model, \( Y_{t+2} \sim A_t \). This approach ensures a consistent estimate with a valid confidence interval. Therefore, we use it as a benchmark for comparing the estimation results of the proposed method. 

\noindent {\bf Estimation Method II: A2-WCLS}. In our simulation study, we applied the A2-WCLS method as specified in Equation \eqref{eq:sim-eq}. The pre-treatment auxiliary variable used for adjustment was \( Z_{t,j} \). Since our primary interest was estimating the fully marginal causal excursion effect (i.e., \( S_t = \emptyset \), \( f_t(S_t) = 1 \)), we set the working model for the centering function under the A2-WCLS criterion in (\ref{eq:a2wcls_working}) as \( \mu_t(S_t) = \theta \), with the corresponding estimate given by $\hat\theta = \frac{\P_n \left[ \sum_{t=1}^T \tilde p_{t,j} (1-\tilde p_{t,j})Z_{t,j}\right]}{\P_n\left[\sum_{t=1}^T \tilde p_{t,j} (1-\tilde p_{t,j})\right]}$. In addition, for post-treatment auxiliary variable adjustment, we used the working model $l_{t+1}(A_{t+1,j}, H_{t+1,j}) = (A_{t+1,j} - p_{t+1,j})(\alpha_{1,1} + Z_{t+1,j}\alpha_{1,2}) + Z_{t+1,j}\alpha_{1,3}$. 

The simulation experiment involves estimating $\theta$ using the data generated, and this introduces non-negligible variation that affects the asymptotic variance estimation for the fully marginal causal effect. Therefore, the asymptotic variance estimator shown in Equation \eqref{eq:adjusted} is used. Table \ref{tab:tabone} reports the simulation results. ``\%RE gain'' indicates the percentage of times we achieve an efficiency gain out of 1000 Monte Carlo replicates. ``mRE'' represents the average asymptotic relative efficiency (i.e., an asymptotic variance ratio), and ``RSD'' represents the relative standard deviation between two estimates. 

\textcolor{black}{Despite having to account for the extra variance caused by estimating $\theta$, the proposed A2-WCLS method still significantly improves the efficiency of fully marginal lagged causal effect estimation. Moreover, as \( \beta_1 \) increases, indicating greater causal model misspecification in the WCLS estimator, A2-WCLS achieves even larger efficiency gains due to its ability to incorporate pre-treatment auxiliary variables that are strong effect moderators, effectively reducing estimation variability.}

\begin{table}[htbp]
\caption{Fully marginal lag $\Delta =2$ causal effect estimation efficiency comparison. The true value of the parameters is $\beta_{0}^\star = -0.1$. \label{tab:tabone}}
\begin{center}
\begin{tabular}{cccccccc}
Method & $\beta_{1}$ & Est & SE & CP & \%RE gain  & mRE & RSD \\\hline
\multirow{3}{*}{WCLS} & 0.2& -0.100 & 0.030  & 0.946 & - & - & -\\
& 0.5&-0.100 & 0.032 & 0.943 & - & - & - \\
&0.8& -0.100 & 0.033  & 0.941 & - & - & - \\
\hline 
\multirow{3}{*}{A2-WCLS} &  0.2& -0.100 & 0.028 & 0.939 & 99.6\% & 1.141 & 1.054\\
& 0.5&-0.100 & 0.029  & 0.937 & 99.8\% & 1.161 & 1.064 \\
&0.8& -0.100 & 0.031 & 0.943 & 100\% & 1.168 & 1.070 \\
\hline
\end{tabular}
\end{center}
\end{table}

\textcolor{black}{Except for estimating the fully marginal lagged effect, we conducted additional simulations to assess the performance of the proposed A2-WCLS estimator. A simulation evaluating the causal excursion effect for the proximal outcome based on Equation \eqref{eq:a2wcls} is provided in Appendix \ref{app:proximal}.
Further evaluations of A2-WCLS in estimating time-varying treatment effects are outlined in Appendix \ref{app:subsec:time-varying}. The results demonstrate consistent estimation and efficiency gains. In Appendix \ref{app:sec:moreTN}, we explore how efficiency gains vary with different choices for sample size $N$ and total time points $T$.}

\section{Application to Intern Health Study}
\label{sec:casestudy}

The Intern Health Study (IHS) is a six-month microrandomized trial on medical interns \citep{necamp2020}, which aimed to investigate when to provide mHealth interventions to people in stressful work environments to improve their behavior and mental health. In this section, we evaluate the effectiveness of targeted notifications in improving an individual's mood and step counts. The data set contains 1562 participants.

The exploratory and MRT analyses conducted in this paper focus on weekly randomization, thus, an individual was randomized to receive mood, activity, sleep, or no notifications with equal probability ($1/4$ each) every week. We choose the outcome $Y_{t+1,j}$ as the average self-reported mood score (a Likert scale taking values from 1 to 10) and step count (cubic root transformed) for individual $j$ in study week $t$. The average weekly mood score when a notification is delivered is 7.14, and 7.16 when there is no notification; The average weekly step count (cubic root) when a notification is delivered is 19.1, and also 19.1 when there is no notification. In the following analysis, the prior week's outcome is chosen as the auxiliary variable. We evaluate the targeted notification treatment effect for medical interns using our proposed method A2-WCLS and WCLS.

\subsection{Comparison of the Marginal Effect Estimation}
\label{sec:casestudy_1}
First, the fully marginal excursion effect is assessed
(i.e., $\beta(t) = \beta_0^\star$). For an individual $j$, the study week is coded as a subscript $t$. $Y_{t+1,j}$ is the self-reported mood score or step count of the individual $j$ in study week $t+1$. $A_{t,j}$ is defined as the specific type of notification that targets improving the outcome. For example, if the outcome is the self-reported mood score, sending mood notifications would be the treatment, thus $\P(A_{t,j}=1) = 0.25$. We analyze the fully marginal causal effect of the targeted notifications on self-reported mood score and step count. 

We considered three estimators for the fully marginal treatment effect $\beta^\star_0$. Estimator I is the unadjusted estimator that uses only the time-varying outcome and treatment assignment data. Estimator II is obtained using the WCLS method \citep{boruvka2018} with the outcome of the previous week $Z_{t,j}:= Y_{t,j}$ as a control variable. Specifically, for mood score outcomes, we selected the previous week's average mood score as the control variable, and for step count outcomes, we chose the prior week's average step count. Estimator III is obtained by further adjusting $Z_{t,j}$ as an auxiliary variable using the proposed A2-WCLS method, with the working centering function defined as $\mu_t=\P_N[Z_{t,j}] =\P_N[Y_{t,j}] $.

We report the estimates in Figure \ref{fig:first} and present more details in Table \ref{tab:fullymarginal} of Appendix \ref{app:casestudy}. Compared to Estimator I, Estimators II \& III have a tangible improvement in standard error estimates, with a relative estimation efficiency of 2.32, 2.33 for the mood outcome and 1.57, 1.58 for the step outcome. Even though Estimator III using A2-WCLS does not gain considerable efficiency over Estimator II (WCLS) results, it is not likely to lose efficiency. We conclude that sending activity notifications can increase (the cubic root of) step counts by 0.072, with statistical significance at level 95\%. However, we do not observe a significant effect of sending mood notifications on users' moods.


\begin{figure}
\begin{center}
\includegraphics[width=5in]{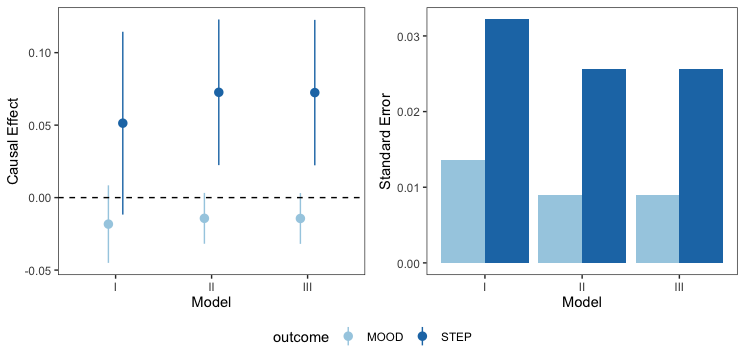}
\end{center}
\caption{Causal effect estimates with the 95\% confidence interval, and standard errors \label{fig:first}}
\end{figure}

\subsection{Time-Varying Treatment Effect Estimation}

In most mobile health intervention studies, time-in-study has been an important moderator of treatment effects due to habituation or learning effect. Therefore, we include study week in the marginal treatment effect model for further investigation: $\beta(t)  = \beta_0^\star+\beta_1^\star t$. The auxiliary variables are still chosen to be the corresponding outcome in the previous week. Control variables have been established in Section \ref{sec:casestudy_1} as being beneficial in reducing standard errors, so control variables (prior week's outcome) will always be included in the following analysis.

Figure \ref{fig:second} below shows estimated time-varying treatment moderation effects.  We compare our proposed approach against the WCLS method from \cite{boruvka2018}. More details of the moderated analysis are presented in Table \ref{tab:timevarying} of Appendix \ref{app:casestudy}. The shaded area represents the 95\% confidence band of the moderation effects at varying values of the moderator (week in study). A narrower confidence band is observed when A2-WCLS is used; specifically, the relative efficiency for $\hat\beta_0, \hat\beta_1$ in the mood model is 2.204 and 2.157, respectively, and in the step model is 1.562 and 1.475, respectively. 


\textcolor{black}{There is an overall decreasing trend in the effectiveness of the treatments on both proximal outcomes with study week. The scientific rationale for this trend may be habituation to or burden caused by the targeted reminders. 
Furthermore, the causal excursion effect of mobile prompts for step count change is positive and statistically significant in the first several weeks of the study. This suggests that sending targeted reminders may be beneficial to increase physical activity in the short term. Note that the estimated effect using A2-WCLS is significant until week 18, while using WCLS the effect is only significant until week 12.  This demonstrates the importance of including auxiliary variables for improved precision in causal moderation analysis.  Both analyses show that in later stages of the study, the effect is no longer significant, which may be due to habituation to and/or burden caused by the smartphone reminders. }

\begin{figure}[htbp]
\begin{center}
\includegraphics[width=5in]{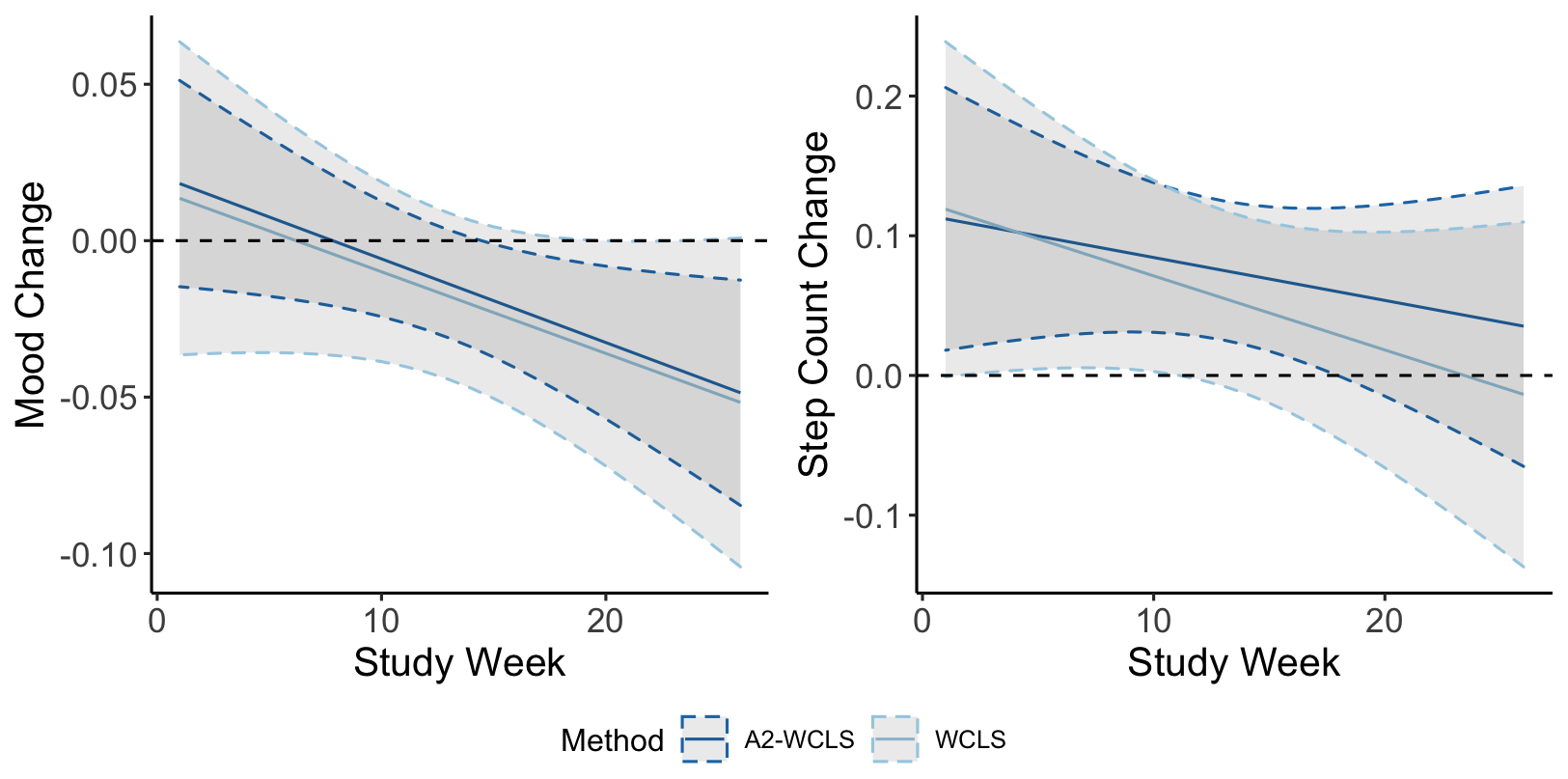}
\end{center}
\caption{Causal effect estimates with the 95\% confidence interval. \label{fig:second}}
\end{figure}

\section{Discussion}
\label{sec:discussion}

\textcolor{black}{MRTs are frequently adopted in mobile health studies to assess the effectiveness of just-in-time intervention components. Although abundant auxiliary variables are often collected in these studies through wearables and smartphones, their statistical utility for enhancing causal effect analysis has not kept pace with its counterpart in classical RCTs. To address this gap, we propose the A2-WCLS approach, which integrates auxiliary variables into the estimation of moderated causal excursion effects. Our work tackles the long-standing questions of (1) how to leverage auxiliary information, (2) when it provides benefits, and (3) the extent of those benefits in the context of repeated measurements and time-varying treatment effects, subsuming existing popular covariate adjustment methods in classical RCTs as a special case (Section \ref{sec:estimand}).}



Under some interpretable conditions in time-varying settings, the proposed A2-WCLS method offers improved efficiency for moderated causal excursion effect estimation in comparison to the benchmark WCLS methods. In Appendix \ref{app:sec:practicalrec}, we provide clear guidelines for practitioners to implement auxiliary variable adjustment in longitudinal studies.

Many open questions remain. First, we may use feature selection methods to regularize the centering function $\mu_t(S_t)$, e.g., by the smooth clipped absolute deviation method (SCAD) \citep{fan2001variable}, to pick an optimal model for $\mu_t(S_t)$. A non-linear centering function can also be considered, similar to \cite{guo2021} proposed for RCTs. In addition to using auxiliary variables to improve estimation efficiency.
Third, if the MRT has a clustered structure (interference within clusters and/or treatment heterogeneity at the cluster level), it requires new centering techniques using cluster-level auxiliary variables \citep{shi2022assessing}. Finally, more advanced and flexible MRT designs should be explored so that they may use auxiliary variable adjustment to maximize efficiency. For example, \cite{van2022} proposes using an information adaptive design, which adapts to the amount of precision gain and can lead to faster, more efficient trials, without sacrificing validity or power. We leave these directions for future work.
\bigskip

\newpage
\appendix

\section{Proof of Lemma \ref{lemma:wcls_u_nonpar}}
\label{app:a2wcls_pertime_wcls}

\subsection{Asymptotic Properties of Nuisance Parameters $\alpha_{0,t}$ and $\alpha_{1,t}$}
\label{app:appendixA1}

Here we first sketch a proof that, given known randomization probability $p_t$ and the centered control variables $\tilde g_t(H_t)^\top$, the time-specific intercepts $\alpha_{0,t}$ and the coefficients $\alpha_{1,t}$ have the same limit value under objective functions displayed in \eqref{eq:unadjusted_nonpar}, \eqref{eq:wcls_nonpar}, and \eqref{eq:a2wcls_pertime}. 
Assume partial derivative and expectation calculations can always be interchangeable. The closed-form estimation for $\alpha_{0,t}$ can be written as:
\begin{align*}
    \alpha^\star_{0,t} &= \E \left[ \left( Y_{t+1}  - \tilde g_t(H_t)^\top \alpha_{1,t} - \left ( A_t -  p_t  \right) (\beta_{0,t} +  \tilde g_t(H_t)^\top \beta_{1,t}) \right) \right] \\
    &= \E \left[ Y_{t+1} \right] - \E \left[ \tilde g_t(H_t)^\top \alpha_{1,t} \right]  - \E \left[ \beta_{0,t}\left ( A_t -  p_t  \right) \right]  - \E \left[ \left ( A_t -  p_t  \right)\left(\tilde g_t(H_t)^\top \beta_{1,t}\right)  \right] \\
    & = \E \left[ Y_{t+1} \right],
\end{align*}
which remains invariant regardless of the covariates adjusted in the regression model.
The rationale lies in both $\E[\tilde g_t(H_t)] = 0$ and $\E[ A_t -  p_t ] = 0$. Besides, we have the limit value for $\alpha_{1,t}$ as:
\begin{align*}
    \alpha^\star_{1,t} &= \Sigma(g_t(H_t))^{-1} \E \left[ \tilde g_t(H_t) \left( Y_{t+1} - \left ( A_t -  p_t  \right) (\beta_{0,t} +  \tilde g_t(H_t)^\top \beta_{1,t}) \right) \right] \\
    &= \Sigma(g_t(H_t))^{-1} \Big\{ \E \left[\tilde g_t(H_t) Y_{t+1} \right]  - \E \left[ \beta_{0,t} \tilde g_t(H_t) \left ( A_t -  p_t  \right) \right]  - \Sigma(g_t(H_t))\E \left[ \left ( A_t -  p_t  \right)\beta_{1,t}  \right] \Big\} \\
    & = \Sigma(g_t(H_t))^{-1}  \E \left[ \tilde g_t(H_t) Y_{t+1} \right].
\end{align*}
As the objective functions presented in \eqref{eq:wcls_nonpar} and \eqref{eq:a2wcls_pertime} both use the same \emph{control variables} $\tilde g_t(H_t)$, the resulting estimates $\hat\alpha_{1,t}$ from different methods are expected to converge to a common value.

\subsection{Asymptotic Variance Comparison Between $\hat\beta_{0,t}^{\text{U}}$ and $\hat\beta_{0,t}^{\text{WCLS}}$}
\label{app:nonparU_WCLS}

For the unadjusted estimator $\hat\beta_{0,t}^{\text{U}}$, the asymptotic variance can be consistently estimated by $Q_t^{-1}\Sigma_t^{\text{U}} Q_t^{-1}$:
\begin{equation}
    \begin{split}
        Q_t &= p_t (1- p_t )\\
        \epsilon_t^{\text{U}} &= Y_{t+1} - \alpha_{0,t} - \left ( A_t -  p_t  \right) \beta_{0,t}  \\
        \Sigma_t^{\text{U}} &= \E\big[\left( \epsilon_t^{\text{U}}(A_t-  p_t )\right)^2\big]
    \end{split}
\end{equation}

For the WCLS estimator $\hat\beta_{0,t}^{\text{WCLS}}$, the asymptotic variance can be consistently estimated by $Q_t^{-1}\Sigma_t^{\text{WCLS}} Q_t^{-1}$:
\begin{equation}
    \begin{split}
        Q_t &=p_t (1- p_t )\\
        \epsilon_t^{\text{WCLS}} &= Y_{t+1} - \alpha_{0,t} - \tilde g_t(H_t)^\top\alpha_{1,t} - \left ( A_t - p_t  \right) \beta_{0,t}  \\
        \Sigma_t^{\text{WCLS}} &= \E\big[\left( \epsilon_t^{\text{WCLS}}(A_t-  p_t )\right)^2\big]
    \end{split}
\end{equation}

For convenience, we center the control variables $g_t(H_t)$ around their means, without loss of generality. Let $\tilde g_t(H_t)$ represent the centered variables. To ensure clarity in the proof, we assume that $\tilde g_t(H_t)^\top\alpha_{1,t}$ does not include the intercept term. Since the outer term $Q_t$ is the same for both asymptotic variance expressions, therefore, we only need to investigate the difference between $\Sigma_t^{\text{WCLS}}$ and $\Sigma_t^{\text{U}}$:
\begin{align*}
    \Sigma_t^{\text{WCLS}} &= \E\big[\left( (\epsilon_t^{\text{U}}- \tilde g_t(H_t)^\top \alpha_{1,t})(A_t- p_t)\right)^2\big] \\
    &= \Sigma_t^{\text{U}} +
    \E\big[\left( \tilde g_t(H_t)^\top \alpha_{1,t}(A_t- p_t )\right)^2\big] - 2 \E\big[ \epsilon_t^{\text{U}}\tilde g_t(H_t)^\top \alpha_{1,t}(A_t- p_t )^2\big]
\end{align*}
The expansion shows that the difference between $\Sigma_t^{\text{WCLS}}$ and $\Sigma_t^{\text{U}}$ is:
\begin{equation}
    \E\big[\left( \tilde g_t(H_t)^\top \alpha_{1,t}(A_t- p_t )\right)^2\big]- 2 \E\big[ \epsilon_t^{\text{U}} \tilde g_t(H_t)^\top \alpha_{1,t}(A_t- p_t )^2\big]
\end{equation}
The first quadratic term is given by:
\begin{align*}
&\E\big[ (A_t - p_t)^2 \alpha_{1,t}^\top \tilde g_t(H_t) \tilde g_t(H_t)^\top \alpha_{1,t} \big] = p_t (1-p_t) \alpha_{1,t}^\top \Sigma(g_t(H_t)) \alpha_{1,t}
\end{align*}
where $\Sigma(g_t(H_t)) = \E\big[\tilde g_t(H_t) \tilde g_t(H_t)^\top\big] \in \mathbb{R}^{d\times d}$ represents the variance-covariance matrix for the control variables $g_t(H_t)$. The second interaction term could be further simplified as:
\begin{align*}
& \E\big[ \epsilon_t^{\text{U}}\tilde g_t(H_t)^\top \alpha_{1,t}(A_t- p_t )^2\big] \\
=&\E  \big[ \left( \E[ \epsilon_t^{\text{U}} | A_t = 1, H_t ] p_t(1-p_t)^2 + \E[ \epsilon_t^{\text{U}} | A_t = 0, H_t] p_t^2  (1-p_t) \right) \tilde g_t(H_t)^\top \alpha_{1,t}\big]\\
=& p_t (1-p_t) \E  \big[ \left( \E[ \epsilon_t^{\text{U}}| A_t = 1, H_t ] (1-p_t) + E[ \epsilon_t^{\text{U}}| A_t = 0, H_t] p_t  \right) \tilde g_t(H_t)^\top \alpha_{1,t} \big]  \\
=& p_t (1-p_t) \Big ( \E  \big[ \E[ \epsilon_t^{\text{U}} | A_t = 1, H_t ] \tilde g_t(H_t)^\top \alpha_{1,t} \big] \\ 
-& p_t \E \big[ (\E [ \epsilon_t^{\text{U}} | A_t = 1, H_t] - \E [ \epsilon_t^{\text{U}} | A_t = 0, H_t] )\tilde g_t(H_t)^\top \alpha_{1,t} \big] \Big). \end{align*}

It rests to study the following terms:
\begin{align*}
&\E  \big[ \E[ \epsilon_t^{\text{U}} | A_t = 1, H_t ] \tilde g_t(H_t)^\top   \big]  \\
= &\E \big[ \E[ Y_{t+1} - (1 - p_t) \beta_{0,t} | A_t = 1, H_t ] \tilde g_t(H_t)^\top  \big]   \\
= &\E \big[ \E[ Y_{t+1} | A_t = 1, H_t] \tilde g_t(H_t)^\top  \big]  \\
= &\E \big[ (\E[ Y_{t+1} | A_t = 1, H_t] - \E[Y_{t+1}|H_t] + \E[Y_{t+1}|H_t]) \tilde g_t(H_t)^\top   \big]  \\
= &\E \big[ (\E[ Y_{t+1} | A_t = 1, H_t] - \E[Y_{t+1}|H_t])  \tilde g_t(H_t)^\top   \big] +\alpha_{1,t}^\top  \Sigma(g_t(H_t)) \\
= &(1-p_t) \beta_{1,t}^\top \Sigma(g_t(H_t))+ \alpha_{1,t}^\top  \Sigma(g_t(H_t)),
\end{align*}
and similarly
\begin{align*}
    \E  \big[ \E[ \epsilon_t^{\text{U}} | A_t = 0, H_t ] \tilde g_t(H_t)^\top   \big] = -p_t \beta_{1,t}^\top \Sigma(g_t(H_t))+ \alpha_{1,t}^\top  \Sigma(g_t(H_t))
\end{align*}
Then combining these two terms together, we have:
\begin{align}
\label{app:eq:diff_wclsu}
    \Sigma_t^{\text{WCLS}}-\Sigma_t^{\text{U}} &= - p_t (1-p_t) \alpha_{1,t}^\top \Sigma(g_t(H_t))  \left(\alpha_{1,t}+2(1-2p_t) \beta_{1,t}\right)
\end{align}

\subsection{More on the Nuisance Parameters $\alpha_{0,t}$ and $\alpha_{1,t}$}

It is natural to consider the estimation of nuisance parameters $\alpha_{0,t}$ and $\alpha_{1,t}$ and their potential optimality in minimizing the asymptotic variances of $\hat\beta_{0,t}^\text{U}$ and $\hat\beta_{0,t}^\text{WCLS}$. To address this, we provide the proof that the estimators of $\alpha_{0,t}$ and $\alpha_{1,t}$ obtained using objective functions \eqref{eq:unadjusted_nonpar} and \eqref{eq:wcls_nonpar} are exactly the optimal value one can get to minimize the asymptotic variance of $\text{Var}(\hat\beta_{0,t}^\text{U})$ and $\text{Var}(\hat\beta_{0,t}^\text{WCLS})$. Here is a sketch of the illustration using the first case:

The asymptotic variance of the unadjusted estimator, which is shown in the previous section, is the following:
\begin{align*}
    \Sigma_t^{\text{U}} = \E\big[\left( \epsilon_t^{\text{U}}(A_t-  p_t )\right)^2\big],
\end{align*}
where $\epsilon_t^{\text{U}} = Y_{t+1} - \alpha_{0,t} - \left ( A_t -  p_t  \right) \beta_{0,t}$. Thus the ``optimal'' $\alpha_{0,t}$ can be obtained:
\begin{align*}
\argmin_{\alpha_{0,t}}\E\big[(A_t-p_t)^2(Y_{t+1} - \alpha_{0,t} - \left ( A_t -  p_t  \right) \beta_{0,t})^2\big],
\end{align*}
which is equivalent to solving:
\begin{equation}
\label{eq:optimal-alpha0}
    \E\big[(A_t-p_t)^2(Y_{t+1} - \alpha_{0,t} - \left ( A_t -  p_t  \right) \beta_{0,t})\big] = 0.
\end{equation}
For the causal parameter $\beta_{0,t}$ to be consistently estimated, it is essential for this condition to be met in conjunction with the following estimating equation:
\begin{equation}
\label{eq:consistent-beta}
    \E\big[(A_t-p_t)(Y_{t+1} - \alpha_{0,t} - \left ( A_t -  p_t  \right) \beta_{0,t})\big] = 0.
\end{equation}
By solving \eqref{eq:optimal-alpha0} and \eqref{eq:consistent-beta} simultaneously, we got $\alpha_{0,t} = \E[Y_{t+1}]$ and $\beta_{0,t} = \E[Y_{t+1}|A_t = 1] -\E[ Y_{t+1}|A_t =0]$, which matches with the limit value of $\alpha_{0,t}$ provided in Appendix \ref{app:appendixA1}. We have the same conclusion for $\alpha_{1,t}$.

\section{Proof of Lemma \ref{theorem:a2wcls_pertime}}
\label{app:a2wcls_pertime}

\subsection{Asymptotic Properties of $\hat\beta_{0,t}^{\text{L}}$}

As per \cite{lin,boruvka2018}, $\hat\beta_{0,t}^{\text{L}}$ obtained by minimizing \eqref{eq:a2wcls_pertime} is consistent and asymptotically normal. We proceed with the proof by centering $g_t(H_t)$ at its mean and denoting it as $\tilde g_t(H_t)$. And the asymptotic variance can be consistently estimated by $Q_t^{-1}\Sigma_t^{\text{L}} Q_t^{-1}$, where
\begin{equation}
    \begin{split}
        Q_t &= p_t(1- p_t )\\
        \epsilon_t^{\text{L}} &= Y_{t+1} - \alpha_{0,t} -  \tilde g_t(H_t)^\top  \alpha_{1,t} - \left ( A_t - p_t  \right) (\beta_{0,t} +  \tilde g_t(H_t) \beta_{1,t}) \\
        \Sigma_t^{\text{L}} &= \E\big[\left( \epsilon_t^{\text{L}}(A_t-  p_t  )\right)^2\big]
    \end{split}
\end{equation}

\subsection{Asymptotic Variance Comparison Between $\hat\beta_{0,t}^{\text{L}}$ and $\hat\beta_{0,t}^{\text{U}}$}

Since the outer term $Q_t$ is the same for both asymptotic variance expressions, therefore, we only need to investigate the difference between $\Sigma_t^{\text{L}}$ and $\Sigma_t^{\text{U}}$. 
\begin{align*}
    \Sigma_t^{\text{L}} &= \E\big[\left( (\epsilon_t^{\text{U}}-\tilde g_t(H_t)^\top  \alpha_{1,t}-(A_t-p_t)\tilde g_t(H_t)^\top  \beta_{1,t})(A_t- p_t)\right)^2\big] \\
    &= \Sigma_t^{\text{U}} +
    \E\big[\left( \tilde g_t(H_t)^\top  \alpha_{1,t}+(A_t-p_t)\tilde g_t(H_t)^\top  \beta_{1,t}\right)^2(A_t- p_t )^2\big] \\
    & - 2 \E\big[ \epsilon_t^{\text{U}}(\tilde g_t(H_t)^\top  \alpha_{1,t}+(A_t-p_t)\tilde g_t(H_t)^\top  \beta_{1,t})(A_t- p_t )^2\big]
\end{align*}
The expansion shows that the difference between $\Sigma_t^{\text{L}}$ and $\Sigma_t^{\text{U}}$ is:
\begin{equation}
\begin{split}
    & \E\big[\left( \tilde g_t(H_t)^\top  \alpha_{1,t}+(A_t-p_t)\tilde g_t(H_t)^\top  \beta_{1,t}\right)^2(A_t- p_t )^2\big] \\
    & - 2 \E\big[ \epsilon_t^{\text{U}}(\tilde g_t(H_t)^\top  \alpha_{1,t}+(A_t-p_t)\tilde g_t(H_t)^\top  \beta_{1,t})(A_t- p_t )^2\big]
\end{split}
\end{equation}
The first quadratic term is given by:
\begin{align*}
&\E\big[\left( \tilde g_t(H_t)^\top  \alpha_{1,t}+(A_t-p_t)\tilde g_t(H_t)^\top  \beta_{1,t}\right)^2(A_t- p_t )^2\big]\\
=& p_t (1-p_t) \big[ \alpha_{1,t}^\top \Sigma(g_t(H_t)) \alpha_{1,t} + 2(1-2p_t)\alpha_{1,t}^\top \Sigma(g_t(H_t)) \beta_{1,t}+ (1-3p_t+3p_t^2)\beta_{1,t}^\top \Sigma(g_t(H_t)) \beta_{1,t}\big]
\end{align*}
The second interaction term could be further simplified as:
\begin{align*}
& \E\big[ \epsilon_t^{\text{U}}(\tilde g_t(H_t)^\top  \alpha_{1,t}+(A_t-p_t)\tilde g_t(H_t)^\top  \beta_{1,t})(A_t- p_t )^2\big] \\
=&\E  \big[ \left( \E[ \epsilon_t^{\text{U}} | A_t = 1, H_t ] p_t(1-p_t)^2 + \E[ \epsilon_t^{\text{U}} | A_t = 0, H_t] p_t^2  (1-p_t) \right) \tilde g_t(H_t)^\top  \alpha_{1,t}\big]\\
& + \E  \big[ \left( \E[ \epsilon_t^{\text{U}} | A_t = 1, H_t ] p_t(1-p_t)^3 - \E[ \epsilon_t^{\text{U}} | A_t = 0, H_t] p_t^3  (1-p_t) \right) \tilde g_t(H_t)^\top  \beta_{1,t}\big]\\
=& p_t (1-p_t) \E  \big[ \left( \E[ \epsilon_t^{\text{U}}| A_t = 1, H_t ] (1-p_t) + \E[ \epsilon_t^{\text{U}}| A_t = 0, H_t] p_t  \right) \tilde g_t(H_t)^\top  \alpha_{1,t} \big]  \\
& + p_t (1-p_t) \E  \big[ \left( \E[ \epsilon_t^{\text{U}}| A_t = 1, H_t ] (1-p_t)^2 - \E[ \epsilon_t^{\text{U}}| A_t = 0, H_t] p^2_t  \right) \tilde g_t(H_t)^\top  \beta_{1,t} \big]  
\end{align*}

Recall:
\begin{align*}
\E  \big[ \E[ \epsilon_t^{\text{U}} | A_t = 1, H_t ] \tilde g_t(H_t)^\top   \big] &= (1-p_t) \beta_{1,t}^\top \Sigma(g_t(H_t))+ \alpha_{1,t}^\top  \Sigma(g_t(H_t))\\
\E  \big[ \E[ \epsilon_t^{\text{U}} | A_t = 0, H_t ] \tilde g_t(H_t)^\top   \big] &= -p_t \beta_{1,t}^\top \Sigma(g_t(H_t))+ \alpha_{1,t}^\top  \Sigma(g_t(H_t))
\end{align*}
Then combing these two terms together, we have $\Sigma_t^{\text{L}}-\Sigma_t^{\text{U}}$ equals to:
\begin{align}
\label{app:eq:diff_lu}
     - p_t (1-p_t) \big[\left(\alpha_{1,t}+(1-2p_t) \beta_{1,t}\right)^\top \Sigma(g_t(H_t)) \left(\alpha_{1,t}+(1-2p_t) \beta_{1,t}\right)+ p_t(1-p_t) \beta_{1,t}^\top\Sigma(g_t(H_t))\beta_{1,t}\big],
\end{align}
which is guaranteed to be non-positive. 

\subsection{Efficiency Comparison Between $\hat\beta_{0,t}^{\text{L}}$ and $\hat\beta_{0,t}^{\text{WCLS}}$}

Simply applying the differences derived above, namely \eqref{app:eq:diff_wclsu} and \eqref{app:eq:diff_lu}, we can easily get the asymptotic variance difference:
\begin{align}
    \Sigma_t^{\text{L}} - \Sigma_t^{\text{WCLS}} =  - p_t (1-p_t) \big[  (1-3p_t + 3p_t^2) \beta_{1,t}^\top\Sigma(g_t(H_t))\beta_{1,t}\big],
\end{align}
which is guaranteed to be non-positive. 

\section{Time-specific Causal Effect Efficiency Comparison with Categorical Moderators}
\label{app:remark3.4}

When discussing the efficiency gain of estimating time-specific \emph{moderated} causal effects, we observe a similar efficiency improvement pattern in obtaining the causal estimator $\beta_{0,t}$ when the causal moderator $f_t(S_t)$ only contains categorical variables. Slightly abusing notations, we redefine the previously mentioned notations as follows. First, for the unadjusted estimator:
\begin{equation}
    \begin{split}
        Q_t &= \E[  p_t(1- p_t ) f_t(S_t)f_t(S_t)^\top]\\
        \epsilon_t^{\text{U}} &= Y_{t+1} - f_t(S_t)^\top\alpha_{0,t} - ( A_t - p_t ) f_t(S_t)^\top\beta_{0,t}  \\
        \Sigma_t^{\text{U}} &= \E\big[   \left( \epsilon_t^{\text{U}}(A_t-  p_t  )\right)^2f_t(S_t)f_t(S_t)^\top\big]
    \end{split}
\end{equation}
Next, for the WCLS estimator:
\begin{equation}
    \begin{split}
        \epsilon_t^{\text{WCLS}} &= Y_{t+1} - f_t(S_t)^\top\alpha_{0,t} -  (Z_t-\mu_t(S_t))^\top \alpha_{1,t} -( A_t - p_t ) f_t(S_t)^\top\beta_{0,t}  \\
        \Sigma_t^{\text{WCLS}} &= \E\big[  \left( \epsilon_t^{\text{WCLS}}(A_t-  p_t  )\right)^2f_t(S_t)f_t(S_t)^\top\big]
    \end{split}
\end{equation}
And, for the \cite{lin} adjusted estimator:
\begin{equation}
    \begin{split}
        \epsilon_t^{\text{L}} &= Y_{t+1} - f_t(S_t)^\top\alpha_{0,t} -  (Z_t-\mu_t(S_t))^\top \alpha_{1,t} - ( A_t - p_t ) \big(f_t(S_t)^\top\beta_{0,t} +  (Z_t-\mu_t(S_t))^\top \beta_{1,t}\big) \\
        \Sigma_t^{\text{L}} &= \E\big[  \left( \epsilon_t^{\text{L}}(A_t-  p_t  )\right)^2f_t(S_t)f_t(S_t)^\top\big] 
    \end{split}
\end{equation}


The intuition behind this lies in the fact that the feature vector $f_t(S_t)$ contains only indicator functions, thus it has only one element that is non-zero with a value of $1$. Furthermore, the product $f_t(S_t)f_t(S_t)^\top$ is a rank-one $q \times q$ diagonal matrix, and there is only one diagonal entry that is $1$. Consequently, the same efficiency comparison holds true for time-specific moderated causal effect estimation. We provide more details in the following. 

\subsection{Asymptotic variance comparison between $\hat\beta_{0,t}^{\text{WCLS}}$ and $\hat\beta_{0,t}^{\text{U}}$}
To simplify the notation, let's assume the auxiliary variable $Z_t$ has a dimension of 1, making $\alpha_{1,t}$ and $\beta_{1,t}$ belong to $\mathbb{R}$. If we have a p-dimensional auxiliary variable (where $p>1$), we can iterate the proof p times to arrive at the same conclusion. 
Since the outer term $Q_t$ is the same for both asymptotic variance expressions, therefore, we only need to investigate the difference between $\Sigma_t^{\text{WCLS}}$ and $\Sigma_t^{\text{U}}$. 
\begin{align*}
    \Sigma_t^{\text{WCLS}} &= \E\big[ \left( (\epsilon_t^{\text{U}}- (Z_t-\mu_t(S_t)) \alpha_{1,t})(A_t- p_t)\right)^2f_t(S_t)f_t(S_t)^\top\big] \\
    &= \Sigma_t^{\text{U}} +
    \E\big[ \left( (Z_t-\mu_t(S_t)) \alpha_{1,t}(A_t- p_t )\right)^2f_t(S_t)f_t(S_t)^\top\big] - \\
    &~~~~ 2 \E\big[  \epsilon_t^{\text{U}}(Z_t-\mu_t(S_t)) \alpha_{1,t}(A_t- p_t )^2f_t(S_t)f_t(S_t)^\top\big]
\end{align*}
The expansion shows that the difference between $\Sigma_t^{\text{WCLS}}$ and $\Sigma_t^{\text{U}}$ is:
\begin{equation}
\begin{split}
        & \E\big[ \left( (Z_t-\mu_t(S_t))\alpha_{1,t}(A_t- p_t )\right)^2f_t(S_t)f_t(S_t)^\top\big]- \\
        & ~~~2 \E\big[  \epsilon_t^{\text{U}} (Z_t-\mu_t(S_t))\alpha_{1,t}(A_t- p_t )^2f_t(S_t)f_t(S_t)^\top\big]
\end{split}
\end{equation}
The first quadratic term is given by:
\begin{equation*}
\E\big[ (A_t - p_t)^2 \alpha_{1,t}^2 (Z_t-\mu_t(S_t))^2 f_t(S_t)f_t(S_t)^\top\big] = p_t (1-p_t)\alpha_{1,t}^2 \Sigma_{\mu}(Z_t)f_t(S_t)f_t(S_t)^\top 
\end{equation*}
where $\Sigma_{\mu}(Z_t) = \E\big[(Z_t-\mu_t(S_t))^2\big] \in \mathbb{R} $. The second interaction term could be further simplified as:
\begin{align*}
& \E\big[  \epsilon_t^{\text{U}}(Z_t-\mu_t(S_t)) \alpha_{1,t}(A_t- p_t )^2f_t(S_t)f_t(S_t)^\top\big] \\
=&\E  \big[  \left( \E[ \epsilon_t^{\text{U}} | A_t = 1,Z_t,S_t] p_t(1-p_t)^2 + \E[ \epsilon_t^{\text{U}} | A_t = 0,Z_t,S_t] p_t^2  (1-p_t) \right)\\
&~~~~~~~~~~~~~~~~~~~~~~~~~~~~~~~~~~~~~~~~~~~~~~~~~~~~~~~~~~~\times (Z_t-\mu_t(S_t))\alpha_{1,t}f_t(S_t)f_t(S_t)^\top\big]\\
=& p_t (1-p_t) \E  \big[  \left( \E[ \epsilon_t^{\text{U}}| A_t = 1,Z_t,S_t ] (1-p_t) + \E[ \epsilon_t^{\text{U}}| A_t = 0,Z_t,S_t] p_t  \right)\\
&~~~~~~~~~~~~~~~~~~~~~~~~~~~~~~~~~~~~~~~~~~~~~~~~~~~~~~~~~~~\times (Z_t-\mu_t(S_t))\alpha_{1,t}f_t(S_t)f_t(S_t)^\top \big] 
\end{align*}

It rests to study the following terms:
\begin{align*}
&\E  \big[ \E[ \epsilon_t^{\text{U}} | A_t = 1, Z_t,S_t ] (Z_t-\mu_t(S_t))\alpha_{1,t}f_t(S_t)f_t(S_t)^\top  \big]  \\
= &\E \big[ \E[ Y_{t+1} | A_t = 1,Z_t,S_t] (Z_t-\mu_t(S_t))f_t(S_t)f_t(S_t)^\top  \big]  \\
= & \E \big[(\E[ Y_{t+1} | A_t = 1, Z_t,S_t] - \E[Y_{t+1}|Z_t,S_t] + \E[Y_{t+1}|Z_t,S_t]) (Z_t-\mu_t(S_t))f_t(S_t)f_t(S_t)^\top  \big]  \\
= &\E \big[ (\E[ Y_{t+1} | A_t = 1, H_t] - \E[Y_{t+1}|Z_t,S_t])  (Z_t-\mu_t(S_t))f_t(S_t)f_t(S_t)^\top  \big] \\
&~~~~~~~~~~~~~~~~~~~~~~~~~~~~~~~~~~~~~~~~~~~~~~~~~~~~~~~~~~~~~~~~~~ +\alpha_{1,t}\Sigma_{\mu}(Z_t)f_t(S_t)f_t(S_t)^\top \\
= &\big((1-p_t) \beta_{1,t}\Sigma_{\mu}(Z_t)+ \alpha_{1,t}  \Sigma_{\mu}(Z_t)\big)f_t(S_t)f_t(S_t)^\top,
\end{align*}
and similarly
\begin{align*}
    \E  \big[ \E[ \epsilon_t^{\text{U}} | A_t = 0, H_t ] (Z_t-\mu_t(S_t))f_t(S_t)f_t(S_t)^\top   \big] = \big(-p_t \beta_{1,t}\Sigma_{\mu}(Z_t)+ \alpha_{1,t}  \Sigma_{\mu}(Z_t)\big)f_t(S_t)f_t(S_t)^\top.
\end{align*}
Then combining these two terms, we have:
\begin{align}
\label{app:eq:diffwclsu_ft}
    \Sigma_t^{\text{WCLS}}-\Sigma_t^{\text{U}} &=
    - p_t (1-p_t)\alpha^2_{1,t} \Sigma_{\mu}(Z_t)  \left(1+2(1-2p_t) \frac{\beta_{1,t}}{\alpha_{1,t}}\right)f_t(S_t)f_t(S_t)^\top 
\end{align}

\subsection{Asymptotic Variance Comparison Between $\hat\beta_{0,t}^{\text{L}}$ and $\hat\beta_{0,t}^{\text{U}}$}

Similarly, we only need to investigate the difference between $\Sigma_t^{\text{L}}$ and $\Sigma_t^{\text{U}}$. 
\begin{align*}
    \Sigma_t^{\text{L}} &= \E\big[\left( (\epsilon_t^{\text{U}}-(Z_t-\mu_t(S_t)) \alpha_{1,t}-(A_t-p_t)(Z_t-\mu_t(S_t))\beta_{1,t})(A_t- p_t)\right)^2f_t(S_t)f_t(S_t)^\top\big] \\
    &= \Sigma_t^{\text{U}} +
    \E\big[\left( (Z_t-\mu_t(S_t)) \alpha_{1,t}+(A_t-p_t)(Z_t-\mu_t(S_t))  \beta_{1,t}\right)^2(A_t- p_t )^2f_t(S_t)f_t(S_t)^\top\big] \\
    & - 2 \E\big[ \epsilon_t^{\text{U}}((Z_t-\mu_t(S_t))^\top  \alpha_{1,t}+(A_t-p_t)(Z_t-\mu_t(S_t))^\top \beta_{1,t})(A_t- p_t )^2f_t(S_t)f_t(S_t)^\top\big]
\end{align*}
The expansion shows that the difference between $\Sigma_t^{\text{L}}$ and $\Sigma_t^{\text{U}}$ is:
\begin{equation}
\begin{split}
    & \E\big[\big( (Z_t-\mu_t(S_t)) \alpha_{1,t}+(A_t-p_t)(Z_t-\mu_t(S_t))  \beta_{1,t}\big)^2(A_t- p_t )^2f_t(S_t)f_t(S_t)^\top\big] \\
    & - 2 \E\big[ \epsilon_t^{\text{U}}\big((Z_t-\mu_t(S_t))\alpha_{1,t}+(A_t-p_t)(Z_t-\mu_t(S_t))\beta_{1,t}\big)(A_t- p_t )^2f_t(S_t)f_t(S_t)^\top\big]
\end{split}
\end{equation}
The first quadratic term is given by:
\begin{align*}
&\E\big[\big( (Z_t-\mu_t(S_t))\alpha_{1,t}+(A_t-p_t)(Z_t-\mu_t(S_t))  \beta_{1,t}\big)^2(A_t- p_t )^2f_t(S_t)f_t(S_t)^\top\big]\\
=& p_t(1- p_t )\big(\alpha^2_{1,t}+ \beta^2_{1,t}(1-3p_t+3p_t^2)+2 \alpha_{1,t}\beta_{1,t}(1-2p_t) \big)\Sigma_{\mu}(Z_t) f_t(S_t)f_t(S_t)^\top\\
=& p_t(1- p_t )\big(\alpha_{1,t}+ \beta_{1,t}(1-2p_t)\big)^2 \Sigma_{\mu}(Z_t) f_t(S_t)f_t(S_t)^\top\\
&~~~~~~~~~~~~~~~~~~~~~~~~~~~~~~~~~~~ +p^2_t(1- p_t )^2\beta^2_{1,t} \Sigma_{\mu}(Z_t) f_t(S_t)f_t(S_t)^\top
\end{align*}
The second interaction term could be further simplified as:
\begin{align*}
& \E\big[ \epsilon_t^{\text{U}}\big((Z_t-\mu_t(S_t))\alpha_{1,t}+(A_t-p_t)(Z_t-\mu_t(S_t))\beta_{1,t}\big)(A_t- p_t )^2f_t(S_t)f_t(S_t)^\top\big] \\
=&\alpha_{1,t}\E  \big[ \left( \E[ \epsilon_t^{\text{U}} | A_t = 1, H_t ] p_t(1-p_t)^2 + \E[ \epsilon_t^{\text{U}} | A_t = 0, H_t] p_t^2  (1-p_t) \right) (Z_t-\mu_t(S_t))  f_t(S_t)f_t(S_t)^\top\big]\\
& +\beta_{1,t} \E  \big[ \left( \E[ \epsilon_t^{\text{U}} | A_t = 1, H_t ] p_t(1-p_t)^3 - \E[ \epsilon_t^{\text{U}} | A_t = 0, H_t] p_t^3  (1-p_t) \right) (Z_t-\mu_t(S_t))f_t(S_t)f_t(S_t)^\top \big]\\
=& \alpha_{1,t} p_t (1-p_t) \E  \big[ \left( \E[ \epsilon_t^{\text{U}}| A_t = 1, H_t ] (1-p_t) + \E[ \epsilon_t^{\text{U}}| A_t = 0, H_t] p_t  \right) (Z_t-\mu_t(S_t)) f_t(S_t)f_t(S_t)^\top \big]  \\
& + \beta_{1,t} p_t (1-p_t) \E  \big[ \left( \E[ \epsilon_t^{\text{U}}| A_t = 1, H_t ] (1-p_t)^2 - \E[ \epsilon_t^{\text{U}}| A_t = 0, H_t] p^2_t  \right) (Z_t-\mu_t(S_t)) f_t(S_t)f_t(S_t)^\top \big]  
\end{align*}
Then combing these two terms together, we have $\Sigma_t^{\text{L}}-\Sigma_t^{\text{U}}$ equals to:
\begin{equation}
\label{app:eq:difflu_ft}
\begin{split}
- p_t (1-p_t) \big[\left(\alpha_{1,t}+(1-2p_t) \beta_{1,t}\right)^2 \Sigma_{\mu}(Z_t) + p_t(1-p_t) \beta_{1,t}^2\Sigma_{\mu}(Z_t)\big]f_t(S_t)f_t(S_t)^\top
\end{split}
\end{equation}
which is guaranteed to be negative semi-definite. 

\subsection{Efficiency Comparison Between $\hat\beta_{0,t}^{\text{L}}$ and $\hat\beta_{0,t}^{\text{WCLS}}$}

Simply applying the differences derived above, namely \eqref{app:eq:diffwclsu_ft} and \eqref{app:eq:difflu_ft}, we can easily get the asymptotic variance difference:
\begin{align}
    \Sigma_t^{\text{L}} - \Sigma_t^{\text{WCLS}} =  - p_t (1-p_t) \big[  (1-3p_t + 3p_t^2) \beta_{1,t}^2\Sigma_{\mu}(Z_t)\big]f_t(S_t)f_t(S_t)^\top,
\end{align}
which is guaranteed to be negative semi-definite.

\section{Illustration and Formulation of Condition \ref{con:orthogonality}}

To illustrate the derivation of the orthogonality condition in detail, we write down the derivatives of \eqref{eq:a2wcls} and (\ref{eq:mrtstandard}) with respect to $\beta_0$. The derivative of (\ref{eq:mrtstandard}) with respect to $\beta_0$ is:
\begin{equation}
\label{app:eq_wclsprime}
    \E\left[\sum_{t=1}^T W_t( Y_{t+1} - g_t (H_t)^\top \alpha - (A_t - \tilde p_t)f_t (S_t)^\top \beta_0)(A_{t} - \tilde p_t) f_t(S_t)\right] 
\end{equation}

The derivative of \eqref{eq:a2wcls} with respect to $\beta_0$ is:
\begin{equation}
\label{app:eq_mawclsprime}
    \E\left[\sum_{t=1}^T W_t( Y_{t+1} - g_t (H_t)^\top \alpha - (A_t - \tilde p_t) ( f_t (S_t)^\top \beta_0 + (Z_t-\mu_t(S_t))^\top \beta_1  ))(A_{t} - \tilde p_t) f_t(S_t)\right]
\end{equation}

The intuition is that if the difference between \eqref{app:eq_wclsprime} and \eqref{app:eq_mawclsprime} is always zero (not depending on the form of $\mu_t(S_t)$), then both of them should have the same minimizer for $\beta_0$:
$$
\E\left[\sum_{t=1}^T W_t(A_{t} - \tilde p_t)^2 f_t(S_t)(Z_t-\mu_t(S_t))^\top\beta_1\right] = \mathbf{0}_{q \times 1}
$$
After calculating the expectation of the term above, we can obtain the form of orthogonality constraint shown in Condition \ref{con:orthogonality}. For each $Z^i_t-\mu^i_t(S_t)$ we wish to include:
$$
\E\left[\sum_{t=1}^T \tilde p_t(1 - \tilde p_t) f_t(S_t)(Z^i_t-\mu^i_t(S_t))^\top\right] = \mathbf{0}_{q \times 1}
$$

\subsection{Neyman Orthogonality}
\label{app:neymanorthogonality}

We are interested in the true value $\beta_0^\star$ of the low-dimensional target parameter $\beta_0$. Therefore in the estimation procedure, $\beta_1$ and $\alpha$ are treated as nuisance parameters.  First, we assume that $\beta_0^\star$ satisfies the moment conditions:
$$
\E[\psi(\beta_0^\star; \beta_1^\star,\alpha^\star)] =0 
$$

where $\psi(\beta_0;\beta_1,\alpha)$ is a known score function. In our case, $\psi(\beta_0;\beta_1,\alpha)$ is exactly the estimating equation for $\beta_0$:
\begin{equation*}
    \sum_{t=1}^T W_t\big[Y_{t+1} - g_t (H_t)^\top \alpha -(A_t - \tilde p_t) ( f_t (S_t)^\top \beta_0 + (Z_t-\mu_t(S_t))^\top \beta_1  )\big](A_t - \tilde p_t)f_t (S_t)
\end{equation*}

Here we further require the Neyman orthogonality condition for the score $\psi$. By the definition given in \cite{chernozhukov2018double}, the Gateaux derivative operator with respect to $\beta_1$ is:
$$
G(\beta_1) = \E\left[\sum_{t=1}^T W_t (A_t - \tilde p_t)^2 f_t(S_t) (Z_t-\mu_t(S_t))^\top\right](\beta_1- \beta_{1,0} )
$$
which should vanish when evaluated at the true parameter values $\beta_{1,0} = \beta_1^\star$, which implies that:
$$
\E\left[\sum_{t=1}^T \tilde p_t(1 - \tilde p_t) f_t(S_t)(Z_t-\mu_t(S_t))^\top\right] = \mathbf{0}_{q \times p}
$$
Similarly, we can obtain the form of orthogonality constraint shown in Condition \ref{con:orthogonality}.

\section{Proof of Lemma \ref{theorem:a2wcls_general}}
\label{app:a2-wcls_asymptotic}

The A2-WCLS criterion is:
\begin{equation*}
     \mathbb{P}_n \left[ \sum_{t=1}^T W_t \times \left( Y_{t+1} - g_t(H_t)^\top \alpha - \left ( A_t - \tilde p_t (1 \mid S_t) \right) (f_t (S_t)^\top \beta_0 + (Z_t - \mu_t(S_t))^\top \beta_1 ) \right)^2 \right]
\end{equation*}
which can be expressed as the following estimating equation:
\begin{align}
    U(\alpha, \beta_0, \beta_1; \mu_t) = \E \Bigg[
 \sum_{t=1}^T W_t( Y_{t+1} - g_t (H_t)^\top \alpha - & (A_t - \tilde p_t) ( f_t (S_t)^\top \beta_0 + (Z_t-\mu_t(S_t))^\top \beta_1  )) \nonumber \\
 & \times \left( \begin{array}{c} g_t(H_t) \\ (A_{t} - \tilde p_t) f_t(S_t) \\
 (A_{t} - \tilde p_t)(Z_t - \mu_t(S_t))\end{array} \right)\Bigg]
\end{align}

Assuming the centering functions $\mu_t(S_t)$ satisfy the orthogonality condition \ref{con:orthogonality}, then we have:
\begin{align*}
    &\E\left[\sum_{t=1}^T W_t( Y_{t+1} - g_t (H_t)^\top \alpha - (A_t - \tilde p_t) ( f_t (S_t)^\top \beta_0 + (Z_t-\mu_t(S_t))^\top \beta_1  ))(A_{t} - \tilde p_t) f_t(S_t)\right]\\
    = &\E\left[\sum_{t=1}^T W_t( Y_{t+1} - g_t (H_t)^\top \alpha - (A_t - \tilde p_t)f_t (S_t)^\top \beta_0)(A_{t} - \tilde p_t) f_t(S_t)\right]\\
    & -\E\left[\sum_{t=1}^T W_t(A_{t} - \tilde p_t)^2 f_t(S_t)(Z_t-\mu_t(S_t))^\top\right]\beta_1 ~~~~ \text{(orthogonality condition)} \\
    = & \E\left[\sum_{t=1}^T W_t( Y_{t+1} - g_t (H_t)^\top \alpha - (A_t - \tilde p_t)f_t (S_t)^\top \beta_0)(A_{t} - \tilde p_t) f_t(S_t)\right]  ~~~~~ \text{(WCLS estimating equation)}
\end{align*}
Based on the asymptotic properties proved in \cite{boruvka2018}, $\hat\beta_0^{\text{A2}}$ is consistent and asymptotically normal, and the robust sandwich variance-covariance estimator is applied here to consistently estimate the asymptotic variance of $\hat\beta_0^{\text{A2}}$, which can be retrieved from the middle block diagonal $(q \times q)$ matrix: 
$$Q^{-1} \Sigma Q^{-1} =  \E \left[\dot U(\alpha, \beta_0, \beta_1; \mu_t) \right]^{-1} \E\left[U(\alpha, \beta_0,\beta_1; \mu_t) \right]^ {\otimes 2} \E \left[\dot U(\alpha, \beta_0, \beta_1; \mu_t) \right]^{-1}$$

Specifically, for the WCLS criterion:
\begin{equation}
\label{app:eq:wclsvar}
    \begin{split}
        Q_{\beta_0} &= \E\left[\sum_{t=1}^T \tilde p_t (1- \tilde p_t )f_t(S_t)f_t(S_t)^\top\right],\\
        \epsilon_t^{\text{WCLS}} &= Y_{t+1} - g_t(H_t)^\top \alpha - \left ( A_t - \tilde p_t  \right) f_t(S_t)^\top \beta_{0}, \\
        \Sigma_{\beta_0}^{\text{WCLS}} &= \E\left[\left( \sum_{t=1}^T W_t \epsilon_t^{\text{WCLS}}(A_t- \tilde p_t  )f_t(S_t)\right)\left( \sum_{t=1}^T W_t \epsilon_t^{\text{WCLS}}(A_t- \tilde p_t  )f_t(S_t)\right)^\top\right].
    \end{split}
\end{equation}

For the A2-WCLS criterion:
\begin{equation}
\label{app:eq:a2wclsvar}
    \begin{split}
        Q_{\beta_0} &= \E\left[\sum_{t=1}^T \tilde p_t (1- \tilde p_t )f_t(S_t)f_t(S_t)^\top\right],\\
        \epsilon_t^{\text{A2}} &= Y_{t+1} - g_t(H_t)^\top \alpha - \left ( A_t - \tilde p_t  \right) (f_t(S_t)^\top \beta_{0} +  (Z_t - \mu_t(S_t))^\top \beta_{1}), \\
        \Sigma_{\beta_0}^{\text{A2}} &= \E\left[\left( \sum_{t=1}^T W_t \epsilon_t^{\text{A2}}(A_t- \tilde p_t  )f_t(S_t)\right)\left( \sum_{t=1}^T W_t \epsilon_t^{\text{A2}}(A_t- \tilde p_t  )f_t(S_t)\right)^\top\right].
    \end{split}
\end{equation}

\section{Time-Lagged Outcomes} 
\label{app:lag-outcome}

\subsection{Illustration of Bias in Post-Treatment Variable Adjustment}
\label{app:lag-outcome-bias}

Naively plugging in post-treatment variables without appropriate adjustment can lead to bias, and we illustrate this point. If we directly plug $Z_{t+u}^\top \alpha_1$ into Equation \eqref{eq:mrtstandard} as a control variable:
\begin{align*}
         \mathbb{P}_n \left[ \sum_{t=1}^{T-\Delta+1} W_{t,\Delta-1} W_t \times \Big( Y_{t,\Delta} - g_t(H_t)^\top \alpha_0 - Z_{t+u}^\top \alpha_1 - \left ( A_t - \tilde p_t (1 \mid S_t) \right) f_t (S_t)^\top \beta_0\Big)^2 \right].
\end{align*}
Then the corresponding estimating equation for $\beta_0$ is:
\begin{equation}
\label{app:eq:naive-wcls-lag}
    \E\left[ \sum_{t=1}^{T-\Delta+1} W_{t,\Delta-1} W_t \times \Big( Y_{t,\Delta} - g_t(H_t)^\top \alpha_0 - Z_{t+u}^\top \alpha_1 - \left ( A_t - \tilde p_t (1 \mid S_t) \right) f_t (S_t)^\top \beta_0\Big)( A_t - \tilde p_t (1 \mid S_t) f_t (S_t) \right].
\end{equation}

By setting the Equation \eqref{app:eq:naive-wcls-lag} to 0, we obtain the following equation:
\begin{align*}
    0 &=  \E\Bigg[\E\left[ \sum_{t=1}^{T-\Delta+1} W_{t,\Delta-1} \tilde p_t( 1 - \tilde p_t) \times \Big( Y_{t,\Delta} - g_t(H_t)^\top \alpha_0 - Z_{t+u}^\top \alpha_1 - \left ( 1 - \tilde p_t  \right) f_t (S_t)^\top \beta_0\Big)  f_t (S_t) |H_t, A_t =1 \right] \\
    &~~~~~~ - \E\left[ \sum_{t=1}^{T-\Delta+1} W_{t,\Delta-1} \tilde p_t( 1 - \tilde p_t)  \times \Big( Y_{t,\Delta} - g_t(H_t)^\top \alpha_0 - Z_{t+u}^\top \alpha_1 +  \tilde p_t  f_t (S_t)^\top \beta_0\Big)  f_t (S_t) |H_t, A_t =0 \right] \Bigg],
\end{align*}
which indicates:
\begin{equation}
    \label{app:eq:wrong_beta}
    \E \Big[\E\left[ W_{t,\Delta-1} (Y_{t,\Delta}- Z_{t+u}^\top \alpha_1 )|A_t =1,H_t \right] - \E\left[ W_{t,\Delta-1} (Y_{t,\Delta}- Z_{t+u}^\top \alpha_1 )|A_t =0,H_t \right]|S_t \Big] = f_t(S_t)^\top \beta_0.
\end{equation}

Recall the modeling assumption we made for lagged outcomes is that 
\begin{align*}
    \E \Big[\E\left[ W_{t,\Delta-1} Y_{t,\Delta}|A_t =1,H_t \right] - \E\left[ W_{t,\Delta-1} Y_{t,\Delta}|A_t =0,H_t \right]|S_t \Big] = f_t(S_t)^\top \beta_0^\star,
\end{align*}
thus, the estimated parameter $\hat \beta_0$ obtained by solving Equation \eqref{app:eq:wrong_beta} is attenuated due to the post-treatment variable $Z_{t+u}$, which is dependent on the treatment allocation $A_t$.

\subsection{Proof of Lemma \ref{lemma:lag-effect}}

Assume $\hat\beta^{A2}_0$ minimizes Equation \eqref{eq:a2-wcls-lag}, then we have:
\begin{align*}
        \mathbb{P}_N \Big[ \sum_{t=1}^{T-\Delta+1} & W_t \Big( Y_{t+\Delta} - \sum_{u=1}^{\Delta-1} \big(l_{t+u}(A_{t+u}, H_{t+u}) - \mu^l_{t+u}(H_{t+u})\big) \\   &~~~~~~~~~~~~~~~~~ -\mu^l_{t}(H_{t}) -\big ( A_t - \tilde p_t (1 \mid S_t) \big)\big(f_t (S_t)^\top \beta_0 + (Z_t - \mu_t(S_t))^\top \beta_1 \big) \Big)^2 \Big].
\end{align*}
Denote $l^\star_{t+u}(A_{t+u}, H_{t+u}) = \E[Y_{t+\Delta}|A_{t+u}, H_{t+u}]$ and $L_t^{\star} = \sum_{u=1}^{\Delta-1} \big(l^\star_{t+u}(A_{t+u}, H_{t+u}) - \mu^l_{t+u}(H_{t+u})\big)$. Worth noticing, $\E[L_t^{\star} |H_{t+1}] = \E[L_t|H_{t+1}] = 0$. The asymptotic properties of the A2-WCLS estimator follow from the expansion:

\begin{align}
\label{app:eq:estimatingeq_R}
    & \P_N \Big[\sum_{t=1}^T W_t \left( Y_{t+\Delta}- L_t - (A_t - \tilde p_t (1 | S_t)) \big(f_t (S_t)^\top \hat\beta^{A2}_0 + (Z_t - \mu_t(S_t))^\top \hat\beta_1 \big) \right)(A_t - \tilde p_t (1 | S_t))f_t (S_t) \Big] \nonumber \\
    & = \P_N \Big[\sum_{t=1}^T W_t \left( Y_{t+\Delta}- L_t^\star -  (A_t - \tilde p_t (1 | S_t)) f_t (S_t)^\top \beta_0^{\star} \right)(A_t - \tilde p_t (1 | S_t))f_t (S_t) \Big] \nonumber \\
    & ~~~~~ - \P_N\Big[\sum_{t=1}^T W_t (A_t - \tilde p_t (1 | S_t))^2 f_t (S_t)f_t (S_t)^\top  \Big](\hat\beta^{(R)}_n-\beta^{\star}) \nonumber \\
    & ~~~~~ + \P_N\Big[\sum_{t=1}^T W_t (A_t - \tilde p_t (1 | S_t))^2 (Z_t - \mu_t(S_t))^\top \hat\beta_1 f_t (S_t)  \Big] \nonumber \\
    & ~~~~~ + \P_N\Big[\sum_{t=1}^T W_t (A_t - \tilde p_t (1 | S_t)) (L_t^\star - L_t) f_t (S_t)  \Big].
\end{align}
By the Weak Law of Large Number (WLLN), we have the following:
\begin{align*}
    &\P_N\Big[\sum_{t=1}^T W_t (A_t - \tilde p_t (1 | S_t))^2 f_t (S_t)f_t (S_t)^\top  \Big]  \overset{P}{\to} \E\Big[\sum_{t=1}^T \tilde p_t (1 | S_t)(1-\tilde p_t (1 | S_t)) f_t (S_t)f_t (S_t)^\top  \Big], \\
    &\P_N\Big[\sum_{t=1}^T W_t (A_t - \tilde p_t (1 | S_t)) (L_t^\star - L_t) f_t (S_t)  \Big] \overset{P}{\to} 0 ~~~~\text{(centering by $\mu^l$).}\\
    & \P_N\Big[\sum_{t=1}^T W_t (A_t - \tilde p_t (1 | S_t))^2 (Z_t - \mu_t(S_t))^\top \hat\beta_1 f_t (S_t)  \Big] \overset{P}{\to} 0 ~~~~\text{(the Orthogonality Condition).}
\end{align*}
The second convergence result holds true for any $L_t$ by the step of centering using $\mu^l$, which leads to $\E[L_t^{\star} |H_{t+1}] = \E[L_t|H_{t+1}] = 0$. Thus we have:
\begin{align*}
    &\E\Big[\sum_{t=1}^T W_t (A_t - \tilde p_t (1 | S_t)) (L_t^\star - L_t) f_t (S_t) \Big] \\
    =& \E\Big[\sum_{t=1}^T \E[ W_t (A_t - \tilde p_t (1 | S_t)) (L_t^\star - L_t) f_t (S_t)|H_{t+1}] \Big] \\
    =& \E\Big[\sum_{t=1}^T  W_t (A_t - \tilde p_t (1 | S_t)) \E[L_t^\star - L_t|H_{t+1}] f_t (S_t) \Big] \\
    =& 0.
\end{align*}

Denote $\tilde Y^{(l)}_{t+\Delta} =Y_{t+\Delta}- L^\star_t$. Therefore, when $n \rightarrow \infty$, after solving \eqref{app:eq:estimatingeq_R}, we obtain the following:
\begin{align*}
    N^{1/2}(\hat\beta_0^{A2} -\beta^{\star}) & = N^{1/2}~ \P_N \Big\{\sum_{t=1}^T ~ \E\Big[\sum_{t=1}^T \tilde p_t (1 | S_t)(1-\tilde p_t (1 | S_t)) f_t (S_t)f_t (S_t)^\top  \Big]^{-1} \times \\
    &~~~~~~~~~~~~~~~~~ W_t \left( \tilde Y^{(L)}_{t+\Delta}- (A_t - \tilde p_t (1 | S_t)) f_t (S_t)^\top \beta^{\star} \right)(A_t - \tilde p_t (1 | S_t))f_t (S_t) \Big\} + o_p(1).
\end{align*}
By the definition of $\beta_0^\star$:
\begin{equation*}
    \E \left[\sum_{t=1}^T W_t \left( \tilde Y^{(l)}_{t+\Delta}- (A_t - \tilde p_t (1 | S_t)) f_t (S_t)^\top \beta_0^{\star} \right)(A_t - \tilde p_t (1 | S_t))f_t (S_t) \right] =0
\end{equation*}
Therefore, under regularity conditions, the estimator $\hat\beta_0^{A2} \overset{P}{\to} \beta_0^{\star}$; that is, $\hat\beta_0^{A2}$ is a consistent estimator of $\beta^{\star}$. The influence function for $\hat\beta_0^{A2}$ is:
\begin{align}
        &\sum_{t=1}^T ~  \E\Big[\sum_{t=1}^T \tilde p_t (1 | S_t)(1-\tilde p_t (1 | S_t)) f_t (S_t)f_t (S_t)^\top  \Big]^{-1} \times \nonumber \\
        & ~~~~~~~~~~~~~~~~~~~~~~~~~W_t \left( \tilde Y^{(l)}_{t+\Delta}- (A_t - \tilde p_t (1 | S_t)) f_t (S_t)^\top \beta^{\star} \right)(A_t - \tilde p_t (1 | S_t))f_t (S_t).
\end{align}
Under moment conditions, we have asymptotic normality with variance given by $Q^{-1} \Sigma_l^{\text{A2}} Q^{-1}$, where
\begin{align*}
    Q & = \E\Big[\sum_{t=1}^T \tilde p_t (1 | S_t)(1-\tilde p_t (1 | S_t)) f_t (S_t)f_t (S_t)^\top  \Big], \\
    \Sigma_l^{\text{A2}} &= \E \Big[ \Big(\sum_{t=1}^T W_t \left( \tilde Y^{(l)}_{t+\Delta}- (A_t - \tilde p_t (1 | S_t)) f_t (S_t)^\top \beta^{\star} \right)(A_t - \tilde p_t (1 | S_t))f_t (S_t) \Big)^2\Big],
\end{align*}
due to space constraints, we use $\E[X^2]$ to denote $\E[X X^\top]$. In conclusion, we establish that the estimator minimizing the R-WCLS criterion $\hat\beta_0^{A2}$ is consistent and asymptotically normal: 
\begin{equation*}
    N^{1/2}(\hat\beta_0^{A2}-\beta^{\star}) \sim \mathcal{N} (0, \Sigma_R).
\end{equation*}

\subsection{A General A2-WCLS for Lagged Outcomes}

When \(\pi_{t+u} \neq p_{t+u}\) for some \(0 < u \leq \Delta-1\), the weights $W_{t, u} = \prod_{s=1}^{u} \pi_{t+s} (A_{t+s} | H_{t+s}) /  p_{t+s} (A_{t+s} | H_{t+s})$. do not simplify to 1, requiring explicit computation for each \(u\). Denote the causal effect model as $\gamma_t(A_t, S_t, Z_t) =\big ( A_t - \tilde p_t (1 \mid S_t) \big)\big(f_t (S_t)^\top \beta_0 + (Z_t - \mu_t(S_t))^\top \beta_1 \big)$. In this case, the objective function \eqref{eq:a2-wcls-lag} can be reformulated as follows:
\begin{equation}
\label{eq:a2-wcls-lag-complex}
\begin{split}
    \mathbb{P}_N \bigg[ &\sum_{t=1}^{T-\Delta+1} W_t \bigg( W_{t,\Delta -1}Y_{t,\Delta} - \sum_{u=1}^{\Delta-1} W_{t,u-1}\Big(W_{t+u}\big(l_{t+u}(A_{t+u}, H_{t+u}) - \gamma_t(A_t, S_t, Z_t)\big) -\\
    &~~~~~~~~~~~\E\big[W_{t+u}\big(l_{t+u}(A_{t+u}, H_{t+u}) - \gamma_t(A_t, S_t, Z_t)\big)|H_{t+u}\big]\Big)- W_{t,\Delta -1}\gamma_t(A_t, S_t, Z_t)  \bigg)^2 \bigg].
\end{split}
\end{equation}

\section{Proof of Theorem \ref{th:efficiency}}
\label{app:varreduction}

\subsection{Proof of Theorem \ref{th:efficiency} 1.}
For simplicity, let $\tilde Y_{t+1} = Y_{t+1} - g_t(H_t)^\top \alpha$, representing the component shared between both estimating equations. This step can also be viewed as the first stage of a two-stage least squares procedure, where the nuisance function space \( g_t(H_t) \in \mathbb{R}^d \) is orthogonal to the treatment effect subspace. 
\begin{align*}
    & \E\Big[\sum_{t=1}^T W_t\big(\tilde Y_{t+1} - (A_t- \tilde p_t) f_t(S_t)^\top \beta_0 \big) (A_t- \tilde p_t) f_t(S_t)\Big] \\
    & = \E\Big[\sum_{t=1}^T \big(W_t(A_t- \tilde p_t)\tilde Y_{t+1} - W_t(A_t- \tilde p_t)^2 f_t(S_t)^\top \beta_0 \big)  f_t(S_t)\Big] \\
    & = \E\Big[\sum_{t=1}^T W_t(A_t- \tilde p_t)\tilde Y_{t+1}   f_t(S_t)\Big] - \E\Big[\sum_{t=1}^T W_t(A_t- \tilde p_t)^2 f_t(S_t)^\top \beta_0 f_t(S_t)\Big] \\
    & = \E\Big[\sum_{t=1}^T \Big(W_t(A_t- \tilde p_t) \tilde Y_{t+1} - \tilde p_t (1- \tilde p_t) f_t(S_t)^\top \beta_0 \Big) f_t(S_t) \Big].
\end{align*}
The estimating equation can now be rewritten to emphasize that the auxiliary information is incorporated to specifically reduce noise in the projection of the outcome onto the treatment effect linear subspace. In our specific cases, we write the estimating equation as follows:
\begin{align*}
    \psi^{\text{A2}}(\beta_0) & = \sum_{t=1}^T \Big(W_t(A_t- \tilde p_t) \tilde Y_{t+1} - \tilde p_t (1- \tilde p_t) \big(f_t(S_t)^\top \beta_0 - (Z_t -\mu_t(S_t) )^\top \beta_1\big)\Big) f_t(S_t) \\
    \psi^{\text{WCLS}}(\beta_0) & = \sum_{t=1}^T \Big(W_t(A_t- \tilde p_t) \tilde Y_{t+1} - \tilde p_t (1- \tilde p_t) f_t(S_t)^\top \beta_0 \Big) f_t(S_t)
\end{align*}
The robust asymptotic variance expressions for both $\hat\beta_0^{\text{WCLS}}$ and $\hat\beta_0^{\text{A2}}$ share the same outer term $Q$, therefore, we only need to investigate the difference between $\Sigma^{\text{A2}}$ and $\Sigma^{\text{WCLS}}$. And $\Sigma^{\text{WCLS}}$ can be expanded as below:
\begin{align*}
    \Sigma^{\text{WCLS}} & = \E\big[\psi^{\text{WCLS}}(\beta_0)\big(\psi^{\text{WCLS}}(\beta_0)\big)^\top\big] \\
    \Sigma^{\text{A2}} & = \E\big[\psi^{\text{A2}}(\beta_0)\big(\psi^{\text{A2}}(\beta_0)\big)^\top\big]
\end{align*}
and their difference is:
\begin{align*}
     & \Sigma^{\text{A2}} - \Sigma^{\text{WCLS}} =  \\
     &-2 \E \Big[\Big(\sum_{t=1}^T \big(W_t(A_t- \tilde p_t) \tilde Y_{t+1} - \tilde p_t (1- \tilde p_t) f_t(S_t)^\top \beta_0 \big) f_t(S_t)\Big) \Big(\sum_{t=1}^T \tilde p_t (1- \tilde p_t)(Z_t -\mu_t(S_t) )^\top \beta_1 f_t(S_t)^\top\Big) \Big] \\
     & + \E \Big[\Big(\sum_{t=1}^T \tilde p_t (1- \tilde p_t)(Z_t -\mu_t(S_t) )^\top \beta_1 f_t(S_t)\Big) \Big( \sum_{t=1}^T \tilde p_t (1- \tilde p_t)(Z_t -\mu_t(S_t) )^\top \beta_1 f_t(S_t)^\top\Big)\Big]
\end{align*}
For the cross term above, we have the folloiwng expansion, when $t \geq t'$:
\begin{align*}
    & \E \Big[\Big( \big(W_t(A_t- \tilde p_t) \tilde Y_{t+1} - \tilde p_t (1- \tilde p_t) f_t(S_t)^\top \beta_0 \big) f_t(S_t)\Big) \Big(\tilde p_{t'} (1- \tilde p_{t'})(Z_{t'} -\mu_{t'}(S_{t'}) )^\top \beta_1 f_{t'}(S_{t'})^\top\Big) \Big] \\
    & = \E \Big[\E\big[ \big(W_t(A_t- \tilde p_t) \tilde Y_{t+1} - \tilde p_t (1- \tilde p_t) f_t(S_t)^\top \beta_0 \big) f_t(S_t)|H_t\big] \tilde p_{t'} (1- \tilde p_{t'})(Z_{t'} -\mu_{t'}(S_{t'}) )^\top \beta_1 f_{t'}(S_{t'})^\top \Big] \\
    & = \E\big[\tilde p_t (1- \tilde p_t)\tilde p_{t'} (1- \tilde p_{t'})(Z_t -\mu_t(S_t) )^\top \beta_1 (Z_{t'} -\mu_{t'}(S_{t'}) )^\top \beta_1f_t(S_t)  f_{t'}(S_{t'})^\top \big],
\end{align*}
and the last equation holds because of the correctly specified causal effect model. Now when $t <t'$:
\begin{align*}
        & \E \Big[\Big( \big(W_t(A_t- \tilde p_t) \tilde Y_{t+1} - \tilde p_t (1- \tilde p_t) f_t(S_t)^\top \beta_0 \big) f_t(S_t)\Big) \Big(\tilde p_{t'} (1- \tilde p_{t'})(Z_{t'} -\mu_{t'}(S_{t'}) )^\top \beta_1 f_{t'}(S_{t'})^\top\Big) \Big] \\
        & = \E \big[ \big(W_t(A_t- \tilde p_t) \tilde Y_{t+1} - \beta(t;H_t)  + \beta(t;H_t) - \tilde p_t (1- \tilde p_t) f_t(S_t)^\top \beta_0 \big) f_t(S_t) \times \\
        &~~~~~~~~~~~~~~~~~~~~~~~~~~~~~~~~~~~~~~~~~~~~~~~~~~~~~~~~~~ \big(\tilde p_{t'} (1- \tilde p_{t'})(Z_{t'} -\mu_{t'}(S_{t'}) )^\top \beta_1 f_{t'}(S_{t'})^\top\big) \big] \\
        & = \E \big[ \big(W_t(A_t- \tilde p_t) \tilde Y_{t+1} - \beta(t;H_t)\big)f_t(S_t) \big(\tilde p_{t'} (1- \tilde p_{t'})(Z_{t'} -\mu_{t'}(S_{t'}) )^\top \beta_1 f_{t'}(S_{t'})^\top\big) \big] \\
        & ~~~~~~~~~ + \E\big[\tilde p_t (1- \tilde p_t)\tilde p_{t'} (1- \tilde p_{t'})(Z_t -\mu_t(S_t) )^\top \beta_1 (Z_{t'} -\mu_{t'}(S_{t'}) )^\top \beta_1f_t(S_t)  f_{t'}(S_{t'})^\top \big]\\
        & = \underbrace{\E \big[ W_t(A_t- \tilde p_t) \tilde Y_{t+1} - \beta(t;H_t)|H_t \big]}_{=0} \E \big[ f_t(S_t) \big(\tilde p_{t'} (1- \tilde p_{t'})(Z_{t'} -\mu_{t'}(S_{t'}) )^\top \beta_1 f_{t'}(S_{t'})^\top\big)|H_t \big] \\
        & ~~~~~~~~~ + \E\big[\tilde p_t (1- \tilde p_t)\tilde p_{t'} (1- \tilde p_{t'})(Z_t -\mu_t(S_t) )^\top \beta_1 (Z_{t'} -\mu_{t'}(S_{t'}) )^\top \beta_1f_t(S_t)  f_{t'}(S_{t'})^\top \big]\\
        & = \E\big[\tilde p_t (1- \tilde p_t)\tilde p_{t'} (1- \tilde p_{t'})(Z_t -\mu_t(S_t) )^\top \beta_1 (Z_{t'} -\mu_{t'}(S_{t'}) )^\top \beta_1f_t(S_t)  f_{t'}(S_{t'})^\top \big]
\end{align*}
By assumption, the residual variation in the outcome (projection), not accounted for by the observed history, is independent of the future observables. Therefore, the difference between two meat terms is:
\begin{align*}
    & \Sigma^{\text{A2}} - \Sigma^{\text{WCLS}}  =\\
    &~~~ -\E \Big[\Big(\sum_{t=1}^T \tilde p_t (1- \tilde p_t)(Z_t -\mu_t(S_t) )^\top \beta_1 f_t(S_t)\Big) \Big( \sum_{t=1}^T \tilde p_t (1- \tilde p_t)(Z_t -\mu_t(S_t) )^\top \beta_1 f_t(S_t)^\top\Big)\Big]
\end{align*}
In conclusion, the estimation efficiency gain is guaranteed.

\subsection{Proof of Theorem \ref{th:efficiency} 2.}

In the main text, we've outlined a connection between Neyman orthogonality and our method, hinting at a connection with semiparametric efficiency theory. In this section, we adopt a semiparametric efficiency theory perspective to show Theorem \ref{th:efficiency} 2 holds based on a semiparametric influence function. A special case arises when $S_t$ is the complete observed history $H_t$, meaning we are interested in a fully conditional causal effect. In such cases, estimators can be obtained using the score function below under reasonable assumptions:
\begin{equation}
\label{eq:effscore}
    S(\psi) =  \sum_{t=1}^{T-\Delta+1} W_{t,\Delta-1}\big(Y_{t+\Delta} - \mu_t^l(H_t) - (A_t - p_t(1|H_t))h_t(H_t)^\top \psi \big) (A_t - p_t(H_t))h_t(H_t) 
\end{equation}
where
\begin{equation}
    \begin{split}
        & h_t(H_t) = (A-p_t)\big(f_t(S_t)^\top \beta_0 + (Z_t - \mu_t(S_t))^\top \beta_1 \big),\\
        & \mu_t(H_t) = \E[W_{t,\Delta-1}Y_{t+\Delta}|H_t], \\
        & W_{t, \Delta-1} = \prod_{s=1}^{u} \pi_{t+s} (A_{t+s} | H_{t+s}) /  p_{t+s} (A_{t+s} | H_{t+s}) 
    \end{split}
\end{equation}
Significant progress has been made by \cite{murphy2001marginal, qian2021} and \cite{shi2022assessing} in the field of semiparametric efficient estimation of the causal excursion effect. It has been demonstrated that an optimal estimating function is orthogonal to the score functions for the treatment selection probabilities \citep{bickel1993efficient}, which means that the estimator can be improved by subtracting the estimating equation from its projection on the score functions for the treatment selection probabilities between \( t+1 \) and \( t+\Delta-1 \) \citep{murphy2001marginal}. If we set $l^\star_{t+u}(A_{t+u}, H_{t+u}) = \E[W_{t+u,\Delta -1-u} Y_{t+\Delta}|A_{t+u}, H_{t+u}],~0<u \leq \Delta-1$. Therefore, the efficient influence function of $\beta_0$ is:
\begin{equation}
\label{eq:semi-eff}
\begin{split}
        S_{\text{eff}} (\psi) = \mathbb{P}_N \bigg[ &\sum_{t=1}^{T-\Delta+1} \bigg( W_{t,\Delta -1}Y_{t+\Delta} - \\
        &~~~~\sum_{u=1}^{\Delta-1} W_{t,u-1}\Big(W_{t+u}\big(l^\star_{t+u}(A_{t+u}, H_{t+u}) - (A_t -  p_t(H_t))h_t(H_t)\big) -\\
    &~~~~~~~~~~~\E\big[W_{t+u}\big(l^\star_{t+u}(A_{t+u}, H_{t+u}) - (A_t -  p_t(H_t))h_t(H_t)\big)|H_{t+u}\big]\Big)- \\
    &~~~~~~~~~~~\mu^{l^\star}_t(H_t) - W_{t,\Delta -1}(A_t -  p_t(H_t))h_t(H_t)  \bigg) (A_t -  p_t(H_t))h_t(H_t)  \bigg].
\end{split}
\end{equation}
Semiparametric efficiency theory establishes that the solution \(\hat{\psi}\) to \(\mathbb{P}_n \left[ S_{\text{eff}} (\psi) \right] = 0\) attains the semiparametric efficiency bound~\citep{newey1990semiparametric}. For a detailed proof of how Equation \eqref{eq:semi-eff} achieves semiparametric efficiency, see \cite{murphy2001marginal}.

\section{Proof of Proposition \ref{lemma:beta_1}}
\label{app:consistency_beta_1}

The relationship between two residuals $\epsilon_t^{\text{A2}}$ and $\epsilon_t^{\text{WCLS}}$ is:
$$
\epsilon_t^{\text{A2}} = \epsilon_t^{\text{WCLS}}- (A_t - \tilde p_t) (Z_t - \mu_t(S_t))^\top \beta_1
$$
Thus, the estimating equation for parameter $\beta_1$ from the A2-WCLS criterion can be written as:
$$
0 = \E \Big[ \sum_{t=1}^T W_t(\epsilon_t^{\text{WCLS}}- (A_t - \tilde p_t) (Z_t - \mu_t(S_t))^\top \beta_1) (A_t - \tilde p_t)(Z_t - \mu_t(S_t)) \Big] 
$$

Solving the equation above, we have:
\begin{equation}
    \label{eq:beta1}
    \beta_1^\star(\mu_t(S_t)) = \E \Big[ \sum_{t=1}^T \tilde p_t (1-\tilde p_t) \Sigma_{\mu}(Z_t) \Big]^{-1} 
    \E \Big[ \sum_{t=1}^T \tilde p_t (1-\tilde p_t) (\beta(t;Z_t, S_t)- \beta(t;S_t)) (Z_t - \mu_t(S_t)) \Big] 
\end{equation}
where $\Sigma_{\mu}(Z_t) = \E [(Z_t - \mu_t(S_t))(Z_t - \mu_t(S_t))^\top]$.
Therefore, if $Z_t$ is in fact not a moderator, then $\beta(t;Z_t, S_t)- \beta(t;S_t) = 0$ for every $t$. Plug in Equation (\ref{eq:beta1}), we have $\beta_1^\star = 0$. When the limiting value satisfies \(\beta_1^\star = 0\), the A2-WCLS estimator for \(\beta_0\) exhibits the same asymptotic behavior as the WCLS estimator. As a result, there is no change in estimation efficiency.



\subsection{Simulation Illustration for Proposition \ref{lemma:beta_1}}

In Proposition \ref{lemma:beta_1}, we highlight that incorporating an auxiliary variable $Z_t$ that does not serve as an effect moderator does not undermine the current efficiency. To illustrate, we conducted a simulation study estimating the fully marginal effect while including a non-effect moderator $Z_t$. The outcomes are summarized in Table \ref{tab:robust} below. 

\begin{table}[phtb]
\caption{Robustness of the proposed A2-WCLS method (with $N=250$ and $T=30$). The true value of the parameters is $\beta_{0}^\star = -0.2$.
\label{tab:robust}}
\begin{center}
\begin{tabular}{ccccccc}
Method &  Est & SE & CP & \%RE gain  & mRE & RSD \\\hline
{WCLS} & -0.200 & 0.024  & 0.941 & - & - & -\\
{A2-WCLS} & -0.200 & 0.024  & 0.941 & 47.3\% & 1.000 & 1.000\\ \hline
\end{tabular}
\end{center}
\end{table}

\section{Estimation of the Centering Function }
\label{app:sec:finite_super}

\subsection{Finite Population vs. Super-Population Based Inference}

The debate over whether to use finite population inference or hypothetical infinite population inference when estimating ATE in RCTs has been extensively discussed \citep{reichardt1999}. Some covariate adjustment methods, such as those proposed by \cite{lin} and \cite{su2021}, use a finite population framework and are shown to perform well asymptotically. In these methods, baseline variables collected before treatment randomization are used as covariates to improve estimation precision, and the only randomness in the data comes from the random treatment assignment. This approach enables the centering parameter $\bar Z$ to be directly plugged into the model, without introducing additional variation to the estimation process. In contrast, \cite{negi2021revisiting} and \cite{ye2022} investigated the asymptotic precision of ATE under the assumption that the subjects were a random sample from an infinite population. The authors suggest ``we cannot assume the data has been centered in advance and $\E[Z] = 0$ without loss of generality'', therefore the extra variance introduced by estimating $\bar Z$ is taken into account.

MRT analysis slightly differs from the discussion above because the auxiliary variables $Z_t$ are collected repeatedly between treatment randomizations. If $Z_{t}$ depends on previous treatments, even in a finite population, estimating $\mu_t(S_t)$ will bring in non-negligible variances that affect the asymptotic variance of the parameters of interest. Therefore, identifying the study population no longer suffices to determine whether extra uncertainty exists. Instead, we should identify whether $Z_{t,j}(\bar A_{t-1,j}) = z_{t,j}$ remains unchanged regardless of how the previous treatment realizations are altered. If so, we can directly center the auxiliary variables $Z_{t}$ with $\mu_t(S_t)$ without adding any additional variation to the estimation process. Otherwise, it is crucial to take into account the extra uncertainty introduced by estimating $\mu_t(S_t)$ to avoid a conservative asymptotic variance estimation.

\subsection{Estimation Methods}
\label{app:sec:superpop_var}

To address the uncertainty in estimating the centering function parameters, we employ a ``stacking estimating equations" approach, which is a type of M-estimator \citep{carroll2006}. The A2-WCLS criterion introduced in \eqref{eq:a2wcls_working} involves two sets of unknown parameters: ${\alpha, \beta_0, \beta_1}$ and ${\theta}$, and the estimating equation of ${\alpha, \beta_0, \beta_1}$ depends on ${\theta}$. Thus, we can jointly estimate these parameters by ``stacking" their estimating equations and solving for all parameters simultaneously. Then standard arguments can be used to show that $\sqrt{n}(\hat\beta_0^{\text{A2}}-\beta^\star_0)$ converges in distribution to a mean zero normal vector with variance-covariance matrix:
\begin{equation}
\label{eq:superpop_var}
\E \left[\dot U(\hat\beta_0;\hat\alpha,\hat\beta_1,\hat\Theta )\right]^{-1} \times \Sigma(\hat\beta_0;\hat\alpha,\hat\beta_1,\hat\Theta) \times \E \left[\dot U(\hat\beta_0;\hat\alpha,\hat\beta_1,\hat\Theta )\right]^{-1},
\end{equation}
where 
\begin{align*}
\Sigma(\hat\beta_0;\hat\alpha,\hat\beta_1,\hat\Theta) = \E \Big[\Big(U_{\beta_0}(\hat\beta_0;\hat \Theta, \hat\beta_1) - U(\hat\Theta)\Big(\E\Big[\frac{\partial U(\Theta)}{\partial \Theta} \big\vert_{\Theta = \hat \Theta} \Big]\Big)^{-1}\E\Big[\frac{\partial U_{\beta_0}(\hat\beta_0;\Theta, \hat\beta_1)}{\partial \Theta}\big\vert_{\Theta = \hat \Theta} \Big]^\top\Big)^ {\ast 2} \Big].
\end{align*}
More details are presented in Section \ref{app:asymptotic}.

\subsection{Super Population Variance Estimation}
\label{app:asymptotic}


In our estimation model, there are two sets of unknown parameters: $\{\alpha,\beta_0,\beta_1\}$ and $\{\Theta\}$, these parameters can be jointly estimated by ``stacking'' their estimating equations and solving for all parameters simultaneously. 
\begin{align*}
    U_{\beta_0}(\beta_0, \Theta) & = 0\\
    U(\Theta) & = 0
\end{align*}
The adjusted bread term could be written as a $2 \times 2$ matrix shown below:
\begin{equation*}
B = 
    \begin{pmatrix}
        \E\Big[\frac{\partial U_{\beta_0}(\beta_0,\hat\Theta)}{\partial \beta_0}\big\vert_{\beta_0 = \hat \beta_0} \Big] & 
        \E\Big[\frac{\partial U_{\beta_0}(\hat\beta_0,\Theta)}{\partial \Theta}\big\vert_{\Theta = \hat \Theta} \Big] & \\
        0& \E\Big[\frac{\partial U(\Theta)}{\partial \Theta} \big\vert_{\Theta = \hat \Theta} \Big]
    \end{pmatrix}
\end{equation*}
and the stacking estimating equations yield the following meat term:
\begin{equation*}
M = 
\begin{pmatrix}
        \E\big[U_{\beta_0}(\beta_0,\hat\Theta)U_{\beta_0}(\beta_0,\hat\Theta)^\top\big] & 
        \E\big[U_{\beta_0}(\beta_0,\hat\Theta)U(\Theta)^\top \big] & \\
        0& \E\big[U(\Theta)U(\Theta)^\top\big]
    \end{pmatrix}
\end{equation*}
The final adjusted sandwich estimator can be obtained by $B^{-1}M(B^{-1})^\top$, and extra the $(1,1)$ element, which by algebra leads to the expression in Lemma 4.5.

\section{Small Sample Size Adjustment}
\label{app:ssa}

The robust sandwich covariance estimator \citep{mancl2001} for the entire variance matrix is given by $Q^{-1} \Lambda Q^{-1}$.  The first term,~$Q$, is given by
\[
 \sum_{j=1}^N \frac{1}{N} D_{j }^\top W_{j } D_{j } 
\]
where $D_{j }$ is the model matrix for individual~$j$, and $W_{j }$ is a diagonal matrix of individual weights.
The middle term~$\Lambda$ is given by
\[
\sum_{i,j=1}^N \frac{1}{N^2} D_{i }^\top W_{i } (I_{i } - H_{i })^{-1}
\epsilon_{i} \epsilon_{j }^\top (I_{j } - H_{j })^{-1} W_{j } D_{j }
\]
where $I_i$ is an identity matrix of correct dimension, $\epsilon_i$ is the individual-specific residual vector and
\[
H_{j} = D_{j}
\left( \sum_{j=1}^N \frac{1}{N} D_{j}^\top W_{j} D_{j} \right)^{-1}
D_{j }^\top W_{j }
\]
From $Q^{-1} \Lambda Q^{-1}$ we extract $\hat{\Sigma}_{\beta}$.

\section{More on Simulations}
\label{app:sim}

\subsection{Simulation Results for Example \ref{example:est}}
\label{app:sec:centerby_mean}

In this simulation, we define the centering function as the global mean of the auxiliary variable $\bar Z = \frac{1}{n\times T}\sum_{j=1}^n \sum_{t=1}^T Z_{t,j}$, representing a one-dimensional centering function. Table \ref{tab:centerby_mean} below shows the results. 


\begin{table}[phtb]
\caption{Fully marginal causal effect estimation efficiency comparison (centering by the mean). The true value of the parameters is $\beta_{0}^\star = -0.2$.\label{tab:centerby_mean}}
\begin{center}
\begin{tabular}{cccccccc}
Method & $\beta_{11}$ & Est & SE & CP & \%RE gain  & mRE & RSD \\\hline
\multirow{3}{*}{WCLS} & 0.2& -0.197 & 0.029  & 0.952 & - & - & -\\
& 0.5&-0.197 & 0.029  & 0.935 & - & - & - \\
&0.8& -0.194 & 0.031  & 0.936 & - & - & - \\
\hline 
\multirow{3}{*}{Mean-Centered WCLS} & 0.2& -0.205 & 0.029  & 0.952 & 23.5\% & 0.992 & 0.995\\
& 0.5&-0.217 & 0.029  & 0.900 & 67.1\% & 1.010 & 1.004 \\
&0.8& -0.227 & 0.030  & 0.855 & 88.9\% & 1.043 & 1.063 \\
\hline
\end{tabular}
\end{center}
\end{table}

\subsection{Proximal Outcome Simulation Result}
\label{app:proximal}

For the case where \( \Delta = 1 \), only pre-treatment variables can be adjusted. To illustrate Theorem 4.3 (1), we conduct a simulation under the strong assumption that \( Z_{t,j} (\bar A_{t-1,j}) = z_{t,j} \). Under this assumption, centering the auxiliary variable does not introduce additional uncertainty. Table \ref{tab:tabfour} presents the simulation results. In this experiment, the fully marginal causal effect, which remains constant over time, is estimated as \( \beta_0^\star = -0.2 \).

\noindent{\bf Estimation Method I: WCLS}. The WCLS method \citep{boruvka2018} is used with $Z_{t,j}$ as a control variable. This method guarantees to produce a consistent estimate with a valid confidence interval. Thus, it will be used as a reference for comparison of the estimation results from the following method. 

\noindent {\bf Estimation Method II: A2-WCLS}. The proposed A2-WCLS method is used with $Z_{t,j}$ adjusted as an auxiliary variable. Since the fully marginal causal excursion effect (i.e. $S_t = \emptyset$, $f_t(S_t) = 1$) is of primary interest, the working model of the centering function using A2-WCLS criterion in (\ref{eq:a2wcls_working}) is $\mu_t(S_t) = \theta$, and $\hat\theta = \frac{\P_N \left[ \sum_{t=1}^T \tilde p_{t,j} (1-\tilde p_{t,j})Z_{t,j}\right]}{\P_N\left[\sum_{t=1}^T \tilde p_{t,j} (1-\tilde p_{t,j})\right]}$.

\begin{table}[phtb]
\caption{Fully marginal causal effect estimation efficiency comparison. \label{tab:tabfour}}
\begin{center}
\begin{tabular}{cccccccc}
Method & $\beta_{11}$ & Est & SE & CP & \%RE gain  & mRE & RSD \\\hline
\multirow{3}{*}{WCLS} & 0.2& -0.200 & 0.029  & 0.955 & - & - & -\\
& 0.5&-0.200 & 0.029  & 0.951 & - & - & - \\
&0.8& -0.199 & 0.029  & 0.947 & - & - & - \\
\hline 
\multirow{3}{*}{A2-WCLS} & 0.2& -0.200 & 0.027  & 0.956 & 100\% & 1.195 & 1.189\\
& 0.5&-0.200 & 0.027  & 0.953 & 100\% & 1.194 & 1.190 \\
&0.8& -0.199 & 0.027  & 0.941 & 100\% & 1.195 & 1.201 \\
\hline
\end{tabular}
\end{center}
\end{table}


\subsection{Time-varying Causal Effect Estimation}
\label{app:subsec:time-varying}

Beyond the simple choice of the fully marginal effect estimation, we here use the proposed method to show the efficiency gain when estimating the time-varying casual effect. Consider an MRT with known randomization probability and the observation vector being a single state variable $Z_t \sim Unif(z_t - 0.01t, z_t + 0.01t)$ at each decision time $t$, and the state expectations are given by $z_t = \E(Z_t|A_{t-1}, H_{t-1})= 0.05t + 0.1 A_{t-1}$. Let
\begin{equation}
\label{eq:time-var-generativemodel}
    Y_{t+1,j} = \big(\beta_{00}  + \beta_{01} t + \delta_{t,j} + \beta_{11} ( Z_{t,j}- \E[Z_{t,j}]) \big)\times \big(A_{t,j} -p_t(1|H_{t,j})\big)+ 0.8 Z_{t,j}  +\epsilon_{t,j} 
\end{equation}
The randomization probability is set to $p_t(1|H_t) = \text{expit}(\eta_1 A_{t-1}+\eta_2 Z_t)$ where $(\eta_1,\eta_2) = (-0.8,  0.8)$ and $\text{expit}(x)=(1+\exp(-x))^{-1}$. 
The error term satisfies $\epsilon_{t} \sim \mathcal{N}(0,1)$ with $\text{Corr}(\epsilon_u, \epsilon_t) = 0.5^{|u-t|/2}$, and the pertubation term $\delta_{t,j} \sim \mathcal{N}(0,1)$. We set $\beta_{00}=-0.2$, $\beta_{01} = 0.02$, and $\beta_{11} = 0.2$ to represent a small moderation effect. Here we assume $\beta(t;S_t) = f_t(S_t)^\top\beta_0^\star = \beta_{00}^\star + \beta_{01}^\star t$, and Table \ref{tab:time-varying} shows the efficiency comparison between the standard WCLS method versus our proposed A2-WCLS. 

\begin{table}[phtb]
\caption{Comparison of efficiency in estimating time-varying treatment effects (with $N=250$ and $T=30$). The true values of the parameters are $\beta_{00}^\star = -0.2$, and $\beta_{01}^\star = 0.02$. \label{tab:time-varying}}
\begin{center}
\begin{tabular}{cccccccc}
Method & Coefficient & Est & SE & CP & \%RE gain  & mRE & RSD \\\hline
\multirow{2}{*}{WCLS} & $\beta_{00}$& -0.199 & 0.063  & 0.948 & - & - & -\\
& $\beta_{01}$& 0.020 & $3.45 \times 10^{-3}$  & 0.948 & - & - & - \\
\hline 
\multirow{2}{*}{A2-WCLS} & $\beta_{00}$& -0.200 & 0.056  & 0.946 & 100\% & 1.262 & 1.257\\
& $\beta_{01}$& 0.020 &  $3.08 \times 10^{-3}$  & 0.951 & 100\% & 1.254 & 1.244 \\
\hline
\end{tabular}
\end{center}
\end{table}


\subsection{Efficiency Gain Magnitude Discussion}
\label{app:sec:moreTN}

In this section, we empirically examine how sample size and the number of time points influence the magnitude of efficiency gains. Table \ref{tab:moreTN} presents results for nine different combinations of sample sizes and time points, where valid and consistent estimates are obtained, as indicated by “mRE” in each cell. The frequency of efficiency gains across \( M = 1000 \) Monte Carlo replications is reported in parentheses. Notably, this frequency increases with larger sample sizes or a greater number of observation time points.

\begin{table}[htbp]
\caption{The average relative efficiency gain across various sample sizes and time points. \label{tab:moreTN}}
\begin{center}
\begin{tabular}{c|ccc}
\hline
mRE & $T = 30$ & $T = 50$ & $T = 100$  \\ \hline
$N = 100$ & $1.171~(98.3\%)$ & $1.172~(98.9\%)$ & $1.168~(99.1\%)$  \\
$N = 250$ & $1.169~(100\%)$ & $1.164~(100\%)$ & $1.168~(100\%)$  \\
$N = 500$ & $1.167~(100\%)$ & $1.166~(100\%)$ & $1.171~(100\%)$  \\
\hline
\end{tabular}
\end{center}
\end{table}

By the Weak Law of Large Numbers, mRE should approximate the true ARE (asymptotic relative efficiency) when we have sufficient replicates (here set to $M=1000$), and the true ARE does not depend on the sample size $N$. This is why we observe consistent average relative efficiency in each column due to extensive replicates (i.e., a large $M$). However, increasing the sample size $N$ does help decrease the variance of the RE estimates, providing a more precise estimate of the true ARE. 

\begin{figure}[htbp]
\begin{center}
\includegraphics[width=6in]{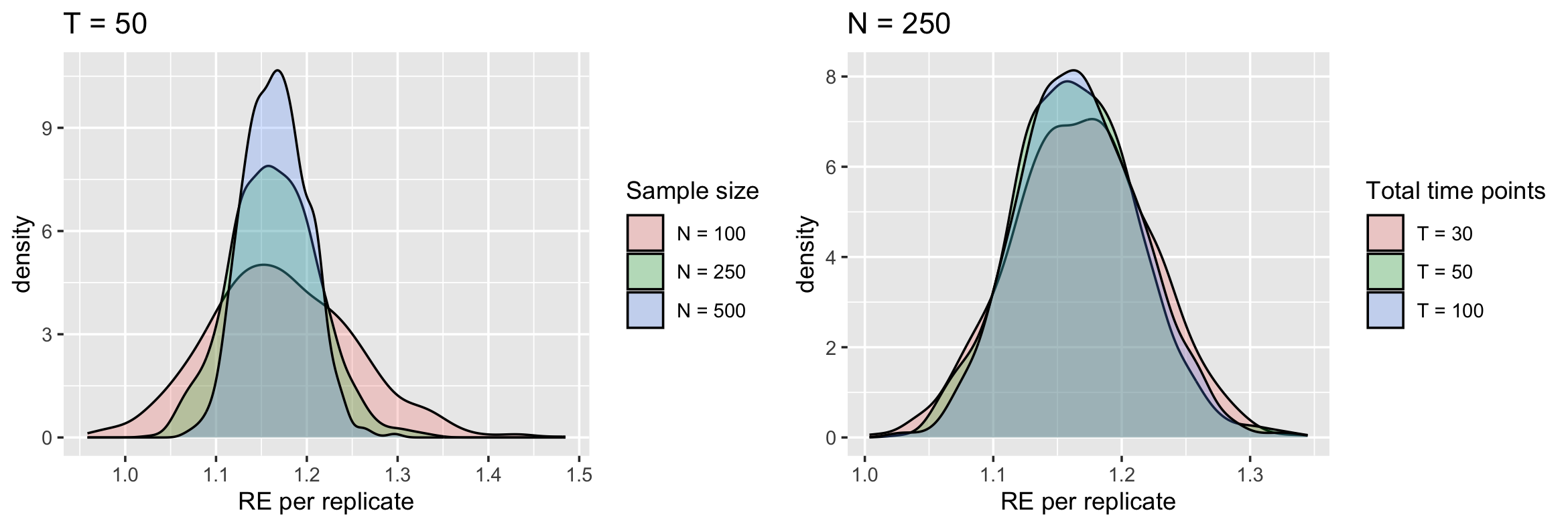}
\end{center}
\caption{Relative efficiency density curve over $M=1000$ replicates. (\textbf{Left}) Fixing total time points $T=50$ and varying sample size $N$.(\textbf{Right}) Fixing sample size $N=250$ and varying total time points $T$. \label{fig:moreTN}}
\end{figure}

Across each row, the distribution of efficiency gains remains relatively stable, showing a slight increase in concentration as \( T \) grows, though not as pronounced as on the left-hand side. This trend can be attributed to the robust variance estimator, which accounts for correlation and effectively ``discounts'' information from additional correlated observations compared to independent ones. Additionally, the extra variance introduced by estimating the centering parameters is also taken into account. Figure \ref{fig:moreTN} shown below depicts how the efficiency magnitude changes when either $N$ or $T$ increases.

\section{Extension: Binary Outcomes}
\label{app:sec_binary}

\subsection{Overview on MRTs with Binary Outcomes}
\label{sec:emeereview}

Many MRTs are interested in binary rather than continuous proximal outcomes. For example, the primary proximal outcome of the Substance Abuse Research Assistance (SARA) study \citep{rabbi2018toward} was whether participants completed the survey and active tasks or not. In BariFit \citep{Barifit}, a study designed to support weight maintenance for individuals who received bariatric surgery, the proximal outcome for the food tracking reminder is whether the participant completes their food log on that day. For binary responses, \cite{qian2021} proposed an estimator of the marginal excursion effect (EMEE) by defining a log relative risk model for the causal excursion effect:
\begin{align}
\label{eq:causalexcursion_binary}
        \beta_{\bf p}(t;s) &= \log \frac{\E \left[ Y_{t+1}(\bar A_{t-1}, A_t = 1)\given S_t(\bar A_{t-1})=s \right]}{\E \left[ Y_{t+1}(\bar A_{t-1}, A_t = 0)\given S_t(\bar A_{t-1})=s \right]}
\end{align}

Similarly, under Assumption~\ref{ass:po},~\eqref{eq:causalexcursion_binary} can be re-expressed in terms of observable data:
\begin{equation}
    \beta_{\bf p}(t;s)= \log \frac{\E \left[ \E \left[Y_{t+1}\given{A_t=1,H_t}\right]\given S_t=s \right]}{\E \left[ \E \left[Y_{t+1}\given{A_t=0,H_t}\right]\given S_t =s\right]}
\end{equation}

Suppose that for $1 \leq t \leq T $, $\beta_b(t;s) = f_t(S_t)^\top \beta_0^{\star \star}$ holds for some q-dimensional parameter $\beta_0^{\star \star}$. We can obtain a consistent estimator by solving the following estimating equation:

\begin{equation}
\label{eq:binaryU}
\mathbb{P}_n \left[ \sum_{t=1}^T W_t e^{-A_{t} f_t (S_t)^\top \beta_0} \left( Y_{t+1} - e^{g_t(H_t)^\top \alpha + A_t f_t (S_t)^\top \beta_0} \right) (A_{t} - \tilde p_t  ) f_t(S_t) \right] = 0
\end{equation}

\subsection{Extensions to Binary Outcomes}

Recall the discussion in Section \ref{sec:emeereview}, it is also possible to apply the proposed methodology to cases where the causal excursion effect is not expressed as a linear contrast, but as logarithmic relative risks \citep{qian2021}. In this section, we will show how auxiliary variables can be incorporated into estimations when the outcome is a binary variable: $Y_{t+1} \in \{0,1 \}$. Similarly, for our auxiliary variable adjusted method to provide us with consistent estimates of the moderated treatment effect, we must ensure orthogonality regarding $\mu_t(S_t)$ and $f_t(S_t)$.


\begin{condition}[\textbf{The orthogonality condition for binary outcomes}]
\label{con:orthogonality_binary}
For a consistent estimation of moderated causal effects, the centering function $\mu_t(S_t)$ must satisfy the following equation:
\begin{equation}
    \E \left[\sum_{t=1}^T W_t e^{-A_t f_t(S_t)^\top \beta_0} Y_{t+1}(1 - e^{-A_t\beta_1 (Z_t-\mu_t(S_t))})(A_t - \tilde p_t)f_t(S_t)\right] = 0
\end{equation}
\end{condition}

It is worth noticing that the orthogonality condition described above is parameter-dependent, i.e., depends on $\beta_0$. An iterative algorithm can be applied to the stacking estimating equations and solve them simultaneously. A simulation study is provided in Section \ref{app:simbinary} to illustrate the efficiency gains.  In the remainder of this paper, we focus exclusively on the linear contrast setting for semi-continuous proximal outcomes.

\subsection{Orthogonality Condition}
The original estimating equation is \citep{qian2021}:
\begin{equation}
\mathbb{P}_n \left[ \sum_{t=1}^T W_t e^{-A_{t} f_t (S_t)^\top \beta_0} \left( Y_{t+1} - e^{g_t(H_t)^\top \alpha + A_t f_t (S_t)^\top \beta_0} \right) (A_{t} - \tilde p_t  ) f_t(S_t) \right] = 0
\end{equation}

After incorporating auxiliary variables, the adjusted estimating equation for $\beta_0$ is:
\begin{equation}
        \mathbb{P}_n \left[ \sum_{t=1}^T W_t  e^{- A_t (f_t(S_t)^\top\beta_0+ \beta_1(Z_{t}-\mu_t(S_t)))} \left( Y_{t+1} - e^{g_t(H_t)^\top \alpha + A_t (f_t (S_t)^\top \beta_0 +\beta_1(Z_{t}-\mu_t(S_t)))} \right) (A_t - \tilde p_t ) f_t(S_t) \right]=0
\end{equation}

To make sure the new estimating equation can still consistently estimate $\beta_0$, the orthogonality condition is:

$$
\E \left[\sum_{t=1}^T W_t e^{-A_t f_t(S_t)^\top \beta_0} Y_{t+1}(1 - e^{-A_t\beta_1 (Z_t-\mu_t(S_t))})(A_t - \tilde p_t)f_t(S_t)\right] = 0
$$

This is a parameter-dependent condition, we can use an iterative algorithm to find a proper $\mu_t(S_t)$. 

\subsection{Simulation Results}
\label{app:simbinary}
Here we implemented a simulation to illustrate the efficiency improvement using the auxiliary variable adjusted method. Table \ref{tab:binaryoutcome} presents the results. 

\begin{table}[phtb]
\caption{Marginal treatment effect estimation efficiency comparison (dynamic treatment probability). \label{tab:binaryoutcome}}
\begin{center}
\begin{tabular}{cccccccc}
Method & $\beta_{11}$ & Bias & SE & CP & \%RE gain  & mRE & RSD \\\hline
\multirow{3}{*}{WCLS} & 0.2& $-2.392 \times 10^{-5}$ & 0.027  & 0.947 & - & - & -\\
& 0.3& $7.078 \times 10^{-4}$ & 0.026  & 0.937 & - & - & - \\
&0.4& $4.787 \times 10^{-4}$ & 0.024  & 0.949 & - & - & - \\
\hline 
\multirow{3}{*}{A2-WCLS} & 0.2& $-6.584 \times 10^{-5}$ & 0.027  & 0.947 & 56.7\% & 1.001 & 0.998\\
& 0.3& $7.643 \times 10^{-4}$ & 0.026  & 0.938 & 51.3\% & 1.000 & 0.998 \\
&0.4& $4.560 \times 10^{-4}$ & 0.024  & 0.949 & 41.1\% & 0.995 & 0.991 \\
\hline
\end{tabular}
\end{center}
\end{table}

\section{More on Case Study}
\label{app:casestudy}

Further details about the estimates are provided in this section. Table \ref{tab:fullymarginal} summarizes the results of using different models to estimate the fully marginal treatment effect. Table \ref{tab:timevarying} compares the estimations between incorporating auxiliary variables versus not, indicating a significant benefit when auxiliary variables are included in estimations. 

\begin{table}[phtb]
\caption{IHS Study: Fully marginal treatment effect estimation. \label{tab:fullymarginal}}
\begin{center}
\begin{tabular}{cccccc}
Outcome & Model & Est & SE & p-value & RE  \\\hline
\multirow{3}{*}{Mood} & I& -0.018 & $1.36 \times 10^{-2}$  & 0.180 & - \\
& II&-0.014 & $8.95 \times 10^{-3}$  & 0.110 & 2.32  \\
&III& -0.014 & $8.94 \times 10^{-3}$  & 0.108 & 2.33 \\
\hline 
\multirow{3}{*}{Steps} & I& 0.051 & $3.22 \times 10^{-2}$  & 0.110 & - \\
& II& 0.073 & $2.56 \times 10^{-2}$  & \textbf{0.005} & 1.57 \\
& III& 0.072 & $2.55 \times 10^{-2}$  & \textbf{0.005} & 1.58 \\
\hline
\end{tabular}
\end{center}
\end{table}

\begin{table}[phtb]
\caption{IHS Study: Time-varying treatment effect estimation. \label{tab:timevarying}}
\begin{center}
\begin{tabular}{ccccccc}
Outcome & Method & Coefficient & Est & SE & p-value & RE  \\\hline
\multirow{4}{*}{Mood} & \multirow{2}{*}{WCLS}& $\hat\beta_0$ & 0.016 & $2.70 \times 10^{-2}$  & 0.551 & - \\
& & $\hat\beta_1$&-0.003 & $1.79 \times 10^{-3}$  & 0.145 & -  \\
& \multirow{2}{*}{A2-WCLS}& $\hat\beta_0$ & 0.021 & $1.79 \times 10^{-2}$  & 0.241 & 2.294 \\
& & $\hat\beta_1$&-0.003 & $1.22 \times 10^{-3}$ & \textbf{0.028} & 2.157 \\
\hline 
\multirow{4}{*}{Steps} & \multirow{2}{*}{WCLS}& $\hat\beta_0$ & 0.124 & $6.48 \times 10^{-2}$  & 0.055 & - \\
& & $\hat\beta_1$&-0.005 & $4.25 \times 10^{-3}$  & 0.212 & -  \\
& \multirow{2}{*}{A2-WCLS}& $\hat\beta_0$ & 0.115 & $5.09 \times 10^{-2}$  & \textbf{0.024} & 1.622 \\
& & $\hat\beta_1$&-0.003 & $3.40 \times 10^{-3}$  & 0.367 & 1.562  \\
\hline
\end{tabular}
\end{center}
\end{table}

\section{Implementation Recommendation}
\label{app:sec:practicalrec}

As a practical recommendation for domain scientists who wish to use A2-WCLS to improve their statistical analysis of time-varying treatment effects, we include this Q\&A session. 

\textbf{\textit{Q1: How to choose the auxiliary variable $Z_t$?}}

Referring to Proposition \ref{lemma:beta_1}, it is technically feasible to incorporate any auxiliary variables that may be relevant. However, practically speaking, doing so is not ideal as it would result in a loss of degrees of freedom and increase the computational burden. We recommend utilizing domain knowledge to identify auxiliary variables that significantly moderate the treatment effect, given existing moderators. This approach allows for a balance between model parsimony and enhanced estimation efficiency.

\textbf{\textit{Q2: How to choose the centering function $\mu_t(S_t)$?}}

Our work has focused on developing a general framework for enhancing the precision of causal effect estimation by incorporating auxiliary variables. This framework can accommodate various choices of the centering function $\mu_t(S_t)$, provided that it satisfies the orthogonality condition. In practice, we recommend using $\mu_t(S_t) = \Theta^\top f_t(S_t) $ as the working model for the centering function, as it is usually low-dimensional and easy to implement, as discussed in Section \ref{sec:properties}. Additionally, our simulation experiments have provided empirical evidence that this linear working model is effective in improving estimation efficiency.

\textbf{\textit{Q3: Does A2-WCLS always lead to an efficiency gain?}}

Assumptions \ref{con:sufficient_1} and \ref{con:sufficient_2} are required for our proposed A2-WCLS method to outperform the benchmark WCLS method. In practice, most MRTs collect a significant amount of information while using relatively simple treatment randomization schemes, making it generally not difficult to satisfy Assumptions \ref{con:sufficient_1} and \ref{con:sufficient_2}. Moreover, in the case where we include an auxiliary variable that is not an effect moderator, we can guarantee that such misspecification would not adversely impact the estimation efficiency. By adopting the A2-WCLS method, we can obtain a consistent and asymptotically more efficient estimator compared to the WCLS method. To account for heteroskedasticity, we use robust standard errors for all estimators. 

\textbf{\textit{Q4: Why should we choose A2-WCLS over WCLS?}}


The proposed A2-WCLS method for incorporating auxiliary variables into causal effect estimation models has been shown to improve or at least maintain estimation precision, as illustrated in both the theoretical and simulation sections. It should be noted that the simulation study conducted using the WCLS approach described in \cite{boruvka2018} may suffer from finite sample bias if strong moderators are not included in the model. In contrast, the proposed A2-WCLS method can alleviate this bias by incorporating strong effect moderators as auxiliary variables in the estimation process, producing a consistent point estimation and a valid variance estimation. Moreover, when assessing lagged causal effects, A2-WCLS offers a structured approach to integrating post-treatment auxiliary variables. Therefore, we recommend the use of the A2-WCLS method to improve the precision of causal effect estimation models.


\section{Code to Replicate Simulation and Case Study Results}
The R code used to generate the simulation experiments and case study results in this paper can be obtained at \verb"https://github.com/Herashi/A2-WCLS".

\newpage
\bibliographystyle{chicago}
\bibliography{paper-ref}

\end{document}